\documentclass[12pt]{article}
\usepackage{tocloft}
\usepackage{amsfonts}
\usepackage{amsmath}
\usepackage{amssymb}
\usepackage{array}
\usepackage{bigints}
\usepackage{bm}
\usepackage{booktabs}
\usepackage[nosort]{cite}
\usepackage{color}
\usepackage{dsfont}
\usepackage{float}
\usepackage{framed}
\usepackage{graphicx}
\usepackage{indentfirst}
\usepackage{mathrsfs}
\usepackage{multirow}
\usepackage{pdflscape}
\usepackage{setspace}
\usepackage{subdepth}
\usepackage{subfig}
\usepackage{titlesec}
\usepackage{wrapfig}
\usepackage[all]{xy}
\usepackage{young}
\usepackage[vcentermath]{youngtab}
\usepackage{relsize}
\usepackage{stackengine}
\usepackage{verbatim}
\usepackage{slashed}

\usepackage{hyperref}
\hypersetup{colorlinks=true}
\hypersetup{linkcolor=black}
\hypersetup{citecolor=black}
\hypersetup{urlcolor=black}

\numberwithin{equation}{section}

\usepackage[left=2.5cm,right=2.5cm,top=2.5cm,bottom=3cm]{geometry}
\fontdimen2\font=1.2\fontdimen2{\jot}{5pt}

\titleformat{\section}{\large\bfseries}{\thesection.}{4pt}{}
\titlespacing{\section}{0pt}{20pt}{6pt}

\titleformat{\subsection}{\normalfont\bfseries}{\thesubsection.}{4pt}{}
\titlespacing{\subsection}{0pt}{15pt}{6pt}

\titleformat{\subsubsection}{\normalfont\itshape}{\thesubsubsection.}{4pt}{}
\titlespacing{\subsubsection}{0pt}{15pt}{6pt}

\titleformat{\paragraph}{\normalfont\itshape}{\theparagraph.}{4pt}{}
\titlespacing{\paragraph}{0pt}{15pt}{6pt}

\def\tilde{\widetilde}
\def\t{\tilde}
\def\hat{\widehat}

\def\bar{\overline}
\def\b{\bar}

\def\half{{1 \over 2}}
\def\d{\partial}

\def\1{{\mathds 1}}

\DeclareMathOperator{\tr}{tr}
\DeclareMathOperator{\Tr}{\mathrm{Tr}}
\DeclareMathAlphabet{\mathbfsf}{OT1}{cmss}{bx}{n}

\newcommand{\Z}{{\mathbb Z}}

\newcommand{\R}{{\mathbb R}}

\def\SL{{\mathscr L}}

\def\CF{{\mathcal F}}

\def\CM{{\mathcal M}}

\def\CO{{\mathcal O}}

\DeclareFontShape{OT1}{cmr}{mx}{n}%
{<->cmr10}{}
\newcommand{\mytitlefont}{\fontseries{mx}\selectfont}
\DeclareMathAlphabet{\titlemath}{OT1}{cmr}{mx}{n}

\begin{document}


\begin{titlepage}
\begin{flushright} \small
UUITP-17/19
 \end{flushright}

\begin{center}
			
~\\[0.7cm]
			
{\fontsize{27pt}{0pt} \mytitlefont Metastable Vacua in Large-$N$ QCD$_3$}
			
~\\[0.4cm]

Adi Armoni,$^{1}$ Thomas T.~Dumitrescu,$^2$  Guido Festuccia,$^{3}$ and Zohar Komargodski\hskip1pt$^{4,5}$

~\\[0.1cm]

$^1$\,{\it Department of Physics, College of Science, Swansea University, SA2 8PP, UK}~\\[0.2cm]

$^2$\,{\it Mani L.\,Bhaumik Institute for Theoretical Physics, Department of Physics and Astronomy,}\\[-5pt]
       {\it University of California, Los Angeles, CA 90095, USA}~\\[0.2cm] 
       
$^3$\,{\it Department of Physics and Astronomy, Uppsala University, SE-75120 Uppsala, Sweden}\\[0.2cm]

$^4$\,{\it Simons Center for Geometry and Physics, SUNY, Stony Brook, NY 11794, USA}\\[0.2cm]

$^5$\,{\it Department of Particle Physics and Astrophysics, Weizmann Institute of Science, Israel}
			
\end{center}

\vskip0.5cm
			
\noindent We reexamine the vacuum structure of three-dimensional quantum chromodynamics (QCD$_3$) with gauge group~$SU(N)$, $N_f$ fundamental quark flavors, and a level-$k$ Chern-Simons term. This analysis can be reliably carried out in the large-$N$, fixed~$N_f, k$ limit of the theory, up to certain assumptions that we spell out explicitly. At leading order in the large-$N$ expansion we find~$N_f + 1$ distinct, exactly degenerate vacuum superselection sectors with different patterns of flavor-symmetry breaking. The associated massless Nambu-Goldstone bosons are generically accompanied by topological Chern-Simons theories. This set of vacua explicitly realizes many candidate phases previously proposed for~QCD$_3$. At subleading order in the large-$N$ expansion, the exact degeneracy between the different superselection sectors is lifted, leading to a multitude of metastable vacua. If we dial the quark masses, different metastable vacua can become the true vacuum of the theory, leading to a sequence of first-order phase transitions.   We show that this intricate large-$N$ dynamics can be captured by the previously proposed bosonic dual theories for QCD$_3$, provided these bosonic duals are furnished with a suitable scalar potential. Interestingly, this potential must include terms beyond quartic order in the scalar fields.

\vfill 
\begin{flushleft} 
May 2019
 \end{flushleft}

\end{titlepage}

	
\setcounter{tocdepth}{3}
\renewcommand{\cfttoctitlefont}{\large\bfseries}
\renewcommand{\cftsecaftersnum}{.}
\renewcommand{\cftsubsecaftersnum}{.}
\renewcommand{\cftsubsubsecaftersnum}{.}
\renewcommand{\cftdotsep}{6}
\renewcommand\contentsname{\centerline{Contents}}
	
\tableofcontents


\section{Introduction}\label{sec:intro}

Gauge theories in three dimensions are qualitatively different from their four-dimensional counterparts. An important reason for this is the existence of the three-dimensional Chern-Simons term,
\begin{equation}\label{CSact}S_\text{CS}={k\over 4\pi} \int d^3x \Tr\left(AdA+{2\over 3}A^3\right)~,
\end{equation}
which defines a well-defined, gauge-invariant contribution to the functional integral as long as the level~$k$ is suitably quantized. Moreover, Chern-Simons terms can arise by integrating out massive fermions, and consequently they are essentially unavoidable~\cite{Niemi:1983rq,Redlich:1983dv}. Since the Chern-Simons action has fewer derivatives than the standard Yang-Mills kinetic term, it is in fact the leading term in the deep infrared, where it gives rise to deconfined low-energy phases with anyons described by a topological quantum field theory (TQFT). The properties of the anyons depend on the gauge group and on the Chern-Simons level~$k$. Another important point is that there is no notion of chirality in three dimensions. Consequently, three-dimensional gauge theories with matter display different symmetries and different patterns of symmetry breaking than their four-dimensional cousins.  

In this paper we will study three-dimensional quantum chromodynamics (QCD$_3$), i.e.~$SU(N)$ Yang-Mills-Chern-Simons gauge theory with gauge group~$SU(N)$ and~$N_f$ Dirac fermions in the fundamental representation (we will follow the notation and conventions in~\cite{Seiberg:2016rsg,Seiberg:2016gmd,Hsin:2016blu,Aharony:2016jvv,Komargodski:2017keh}). This theory has a global~$U(N_f)$ flavor symmetry. The dynamics of the theory depends on the number of colors~$N$, the Chern-Simons level~$k$ in~\eqref{CSact}, the number~$N_f$ of fermion flavors, and the fermion masses. Another mass scale is furnished by the Yang-Mills gauge coupling~$g^2$, which plays a role analogous to that of the strong-coupling scale in four-dimensional asymptotically free gauge theories. 

Several corners of the QCD$_3$ parameter space admit weak-coupling expansions and are consequently rather well understood:

\begin{itemize} 

\item {\it Large masses:} If the mass~$m$ of any fundamental fermion is sufficiently large, we can reliably integrate out that fermion. At one loop, this leads to a well-known shift of the Chern-Simons level~\cite{Niemi:1983rq,Redlich:1983dv}, 
\begin{equation}k \quad \longrightarrow \quad k+\half\, {\rm sgn}(m)~.
\end{equation}
This formula holds for any fundamental quark. If we add the same mass~$m$ for all~$N_f$ quark flavors, we find that the deep IR is described by a Chern-Simons theory with level~$k_\text{IR}=k+{N_f\over 2}$ when~$m$ is large and positive. When~$m$ is large and negative, we instead find~$k_\text{IR}=k-{N_f\over 2}$. The low-energy description of these two regimes consists of~$SU(N)_{k_\text{IR}}$ Chern-Simons TQFTs with two distinct levels~$k_\text{IR}=k\pm{N_f\over 2}$. There is therefore necessarily a phase transition as we vary~$m$ from large negative to positive masses. (This holds for all values of~$k$ and~$N$, as long as~$N_f \geq 1$.) Such a phase transition may be second order, i.e.~described by a Conformal Field Theory (CFT), or it may be first order. More generally, there may be several transitions of different orders as we dial~$m$. 

\newpage

\item {\it Large $k$ or large $N_f$:} In these limits one can solve the model explicitly in perturbation theory (see~\cite{Appelquist:1986fd} for some early work) and one finds a weakly-coupled CFT which describes a second-order transition between the two large-mass phases reviewed in the previous bullet point. 

\item {\it Large $k$ and large $N$ with fixed~$\lambda = N/k$:} The full Yang-Mills-Chern-Simons theory is not solvable in this limit. However, if one discards the Yang-Mills term before taking the limit, one finds a tractable CFT describing a second-order transition between the two large-mass phases described in the first bullet point above. The properties of this CFT can be understood in great detail as a function of~$\lambda$~\cite{Giombi:2011kc ,Aharony:2011jz,Aharony:2012nh,GurAri:2012is,Aharony:2012ns,Jain:2013gza,Jain:2013py,Inbasekar:2015tsa}. See also the review~\cite{Wadia:2016zpd}. In the recent literature this limit has been termed the three-dimensional 't~Hooft limit, with~$\lambda$ the corresponding 't Hooft coupling. We emphasize that this limit is distinct from the conventional 't~Hooft planar, large-$N$ limit in four dimensions~\cite{tHooft:1973alw}, whose three-dimensional analogue is the main subject of this paper. 
\end{itemize} 

Duality plays a crucial role in elucidating the dynamics of QCD$_3$ (see~\cite{Peskin:1977kp,Dasgupta:1981zz,Son:2015xqa,Metlitski:2015eka,Radicevic:2015yla,Aharony:2015mjs, Karch:2016sxi, Murugan:2016zal, Seiberg:2016gmd, Hsin:2016blu, Metlitski:2016dht, Aharony:2016jvv, Benini:2017dus, Armoni:2017jkl} for an incomplete sampling of the original literature). For an earlier work on mass gap in QCD$_3$ see~\cite{Agarwal:2012bn, Agarwal:2015fda}.\footnote{~For some other recent work on non-supersymmetric gauge dynamics in three dimensions see~\cite{Mross:2015idy, Kachru:2016rui, Kachru:2016aon, Karch:2016aux, Wang:2017txt, Jensen:2017dso, Chester:2017vdh, Gomis:2017ixy, Cordova:2017vab, Benini:2017aed,  Jensen:2017bjo, Cordova:2017kue, Chen:2018vmz, Aitken:2018cvh, Armoni:2018ahv, Choi:2018tuh, Sachdev:2018nbk, Cordova:2018qvg,  Benvenuti:2019ujm, DiPietro:2019hqe, Chatterjee:2019xgs}.} Motivated by the idea that the large-mass limits are separated by a single phase transition, which is known to be of second order in the weak-coupling limits reviewed above, it was proposed that the bosonic theory 
\begin{equation}\label{eq:introdual}
U\left(\t N = k+{N_f \over 2}\right)_{-N}+N_f~ \text{fundamental scalars} ~\phi~,
\end{equation}
describes precisely the same phase transition. To see this, assume that all the scalars~$\phi$ are massive 
and can be integrated out. Then the low energy theory is described by the TQFT $U\left(k+{N_f \over 2}\right)_{-N}$, which is level-rank dual to $SU(N)_{k+{N_f \over 2}}$. This in turn is the TQFT describing the large positive quark mass phase of QCD$_3$. 

Other phases can arise if some of the scalars condense. A priori there are many possible condensates, leading to a variety of low-energy theories. In order to reproduce the large negative mass phase of QCD$_3$, one must assume that the scalar potential of the bosonic dual theory is chosen such that the~$\phi$ condensate does not break the~$U(N_f)$ flavor symmetry. This is only possible if the rank~$\t N = k + {N_f \over 2}$ of the dual~$U(\t N)$ gauge theory is bigger than~$N_f$, or alternatively if~$k \geq {N_f \over 2}$. (See section~\ref{sec:duality} for a detailed review.) The fact that the flavor symmetry is unbroken forces the gauge symmetry to be higgsed to~$U\left(k-{N_f\over 2}\right)$. We will occasionally refer to higgsing while preserving the largest possible flavor symmetry as full color-flavor locking. Here this phenomenon leads to a low-energy~$U\left(k-{N_f\over 2}\right)_{-N}$ Chern-Simons TQFT, which is level-rank dual to $SU(N)_{k-{N_f\over 2}}$ and therefore describes the large negative mass phase of QCD$_3$. 

\begin{figure}[!th]
\centerline{\includegraphics[scale=0.42]{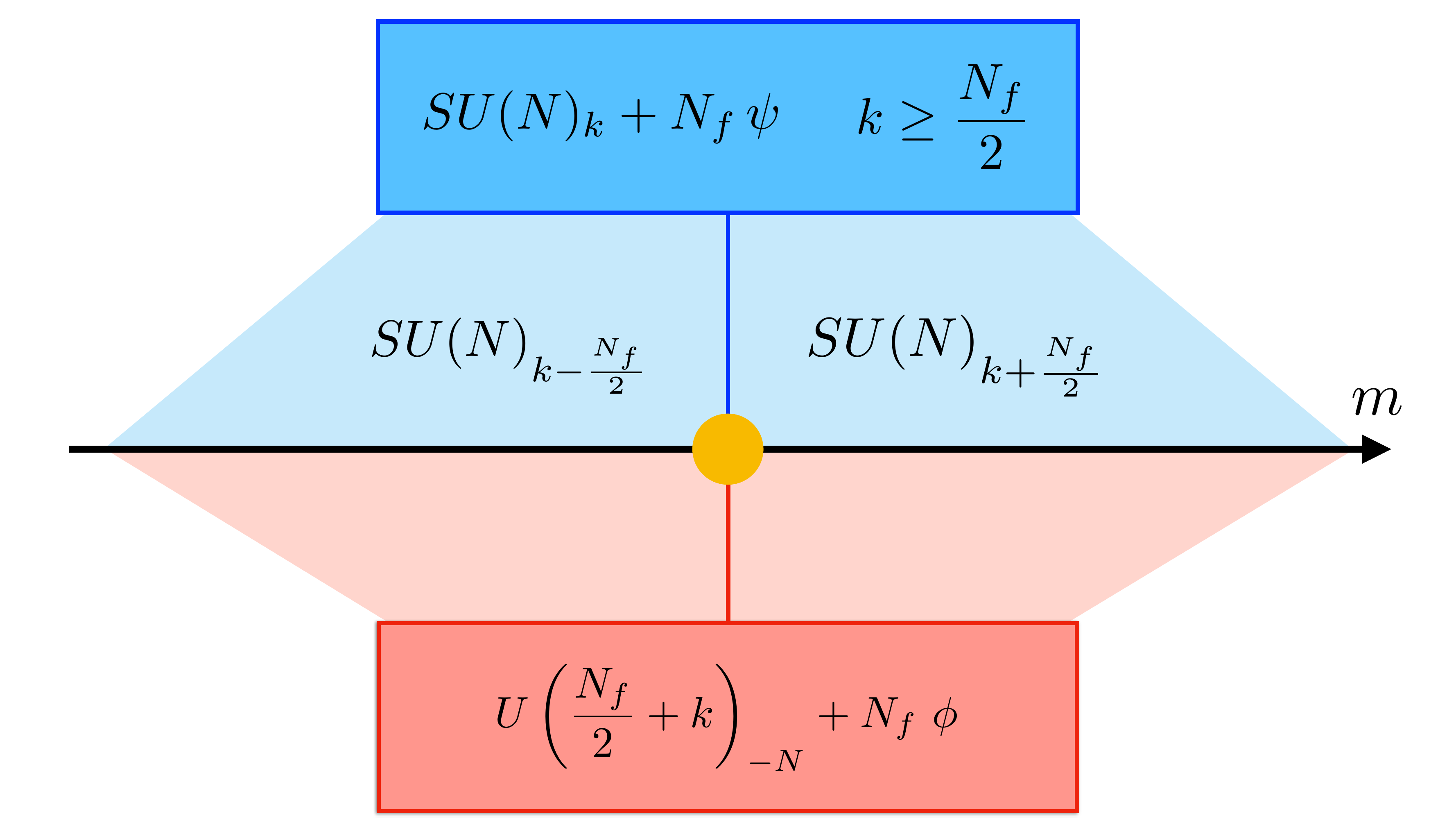}}
\caption{Minimal possible phase diagram for QCD$_3$ (shown in blue) when~$k \geq {N_f \over 2}$. The two massive phases contain non-trivial TQFTs in the deep IR. The bosonic dual is shown in red, and the transition between the two phases (which may be first or second order) is indicated by a yellow dot. In some weak-coupling limits this phase diagram can be established rigorously, and the phase transition can be shown to be second order.}
\label{BasicDiagram}
\end{figure}

These considerations motivate a simple phase diagram for QCD$_3$ with~$k \geq {N_f \over 2}$, which is shown in figure~\ref{BasicDiagram}. Although this phase diagram has been subjected to a multitude of stringent tests, it ultimately remains a conjecture. Moreover, even if the structure of the phases is correct, it is not in general possible to determine the order of the phase transition that separates them. An exception occurs in the weak-coupling limits described in some of the bullet points above, where it can be shown that the transition is second-order and described by an interacting CFT. Whenever the transition is second-order, and hence described by a CFT, the duality implies that both QCD$_3$ and the bosonic dual theory~\eqref{eq:introdual} can be tuned to flow to this CFT in the deep IR.

\begin{figure}[!th]
\centerline{\includegraphics[scale=0.42]{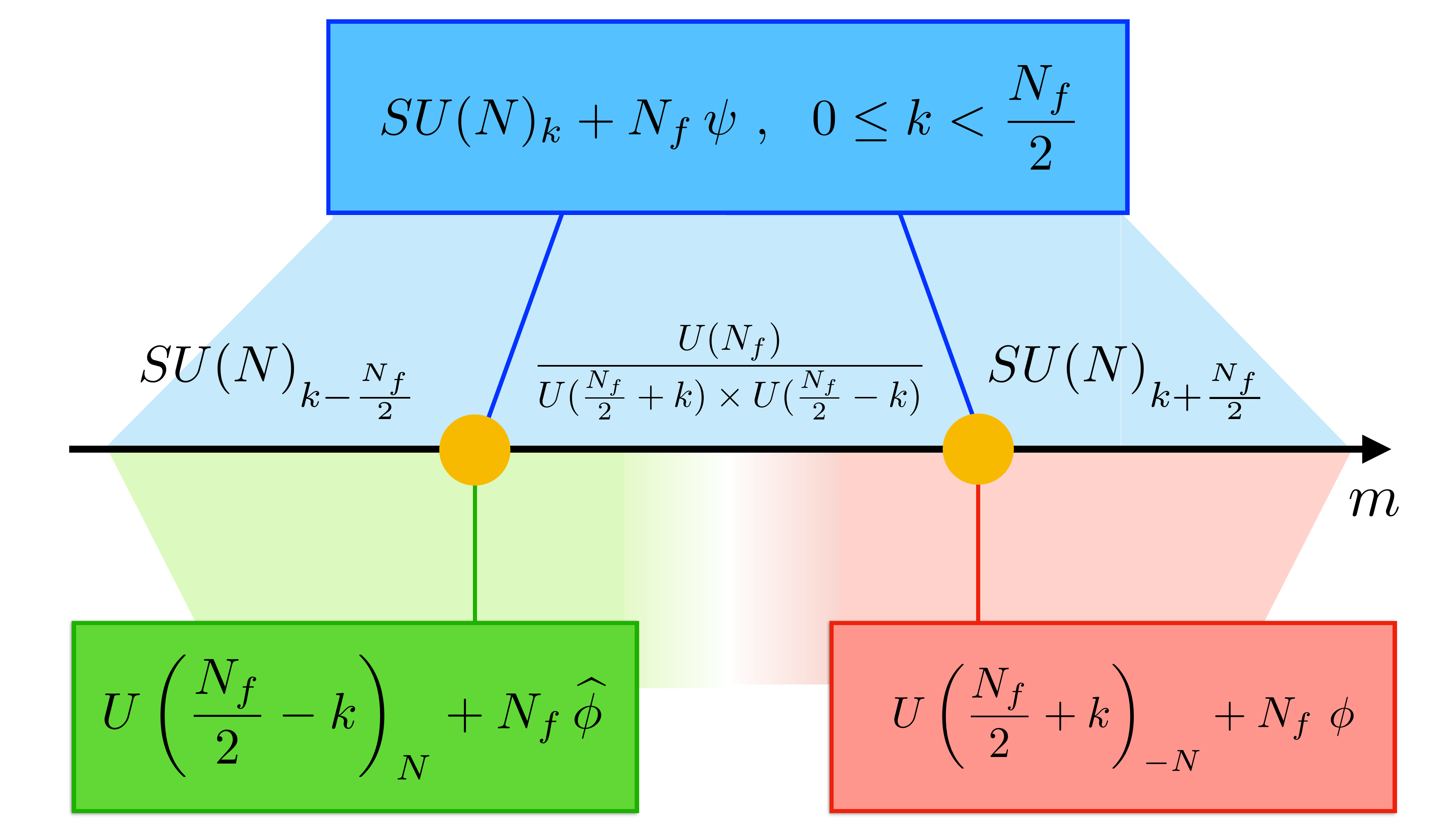}}
\caption{Minimal proposed phase diagram of QCD$_3$ with~$0 \leq k < {N_f \over 2}$ (shown in blue). In addition to the large-mass phases containing Chern-Simons TQFTs, there is an intermediate quantum phase containing NG bosons, but no TQFT. The two phase transitions (which may be first or second order) are indicated by yellow dots. The quantum phase can be simultaneously described by both bosonic dual theories (shown in green and red).}
\label{TriangleD}
\end{figure}

The scenario discussed above, where the condensate in the Higgs phase preserves the full~$U(N_f)$ flavor symmetry, is only possible when~$k \geq {N_f \over 2}$. For this reason, the dynamics of QCD$_3$ in the regime~$0 \leq k < {N_f \over 2}$ is necessarily more involved.\footnote{~As we will discuss below, time reversal flips the sign of $k$ so that we are free to choose $k\geq 0$ without loss of generality.} Note that this regime is beyond any of the weak-coupling limits reviewed above, as long as~$N_f$ is not too large and the quark masses are sufficiently small.\footnote{~More precisely, the phase diagram discussed here is only expected to exist for sufficiently small~$N_f$~\cite{Komargodski:2017keh}. See \cite{Sharon:2018apk} for some rigorous bounds based on the~$F$-theorem.} A minimal possible conjecture~\cite{Komargodski:2017keh} for the behavior of QCD$_3$ when~$0 \leq k < {N_f \over 2}$ is that there are three phases, which can be described by two mutually non-local bosonic duals,\footnote{~Here we have in mind a picture similar to the mutually non-local monopole and dyon points in Seiberg-Witten theory~\cite{Seiberg:1994rs}.} 
\begin{equation}\label{TwoDuals}U\left({N_f\over 2} + k\right)_{-N}+N_f \;\text{fundamentals} \; \;\phi~,\quad U\left({N_f\over 2}-k\right)_{N}+N_f\; \text{fundamentals}\;\; \widehat \phi~.\end{equation}
When the scalars~$\phi$ or~$\widehat \phi$ condense, they necessarily break the~$U(N_f)$ flavor symmetry. However, it is natural to retain the assumption of full color-flavor locking, i.e.~the condensates preserve the largest possible flavor symmetry. With this assumption, both bosonic dual theories in~\eqref{TwoDuals} possess a phase in which the gauge symmetry is completely higgsed, but the flavor symmetry is spontaneously broken as follows, 
\begin{equation}\label{SymmB}U(N_f) \quad \longrightarrow \quad U\left({N_f \over 2} +k\right) \times U\left({N_f \over 2} -k\right)~.\end{equation}
The deep IR contains the corresponding Nambu-Goldstone (NG) bosons, whose target space is the complex Grassmannian~$\text{Gr}\left({N_f \over 2} + k, N_f\right) = U(N_f) \Big/ U\left({N_f \over 2} + k \right) \times U\left({N_f \over 2} - k\right)$. These considerations motivate a minimal phase diagram for QCD$_3$ with~$0 \leq k < {N_f \over 2}$, which is shown in figure~\ref{TriangleD}. As before, this conjectured phase diagram has been subjected to a large number of consistency checks. However, the order of the two  proposed phase transitions is not easy to determine. 

In this paper we will study QCD$_3$, as well as its bosonic duals~\eqref{eq:introdual}, \eqref{TwoDuals} in a limit that is different from any of the weak-coupling limits mentioned above. The limit we consider here is the large-$N$ limit where~$k$, $N_f$ are held fixed, while the Yang-Mills gauge coupling~$g^2$ scales such that the mass scale~$\Lambda = g^2 N$ is held fixed. This limit is simply the three-dimensional version of the standard planar large-$N$ limit introduced by 't Hooft in four dimensions~\cite{tHooft:1973alw}. This limit is qualitatively different from what has come to be known as the 't Hooft limit in three dimensions, where both~$N$ and~$k$ are taken to infinity, but the dimensionless quantity~$\lambda = {N \over k}$ is held fixed (see above). In order to avoid any confusion we refer to the limit we study here (with~$N \rightarrow \infty$ and~$k, N_f$ fixed) as the large-$N$ limit. 

The large-$N$ limit of QCD$_3$ is non-trivial and exhibits new interesting phases and phase transitions. Nevertheless, these phases can be analyzed very explicitly by adapting the logic of~\cite{Coleman:1980mx} to QCD$_3$. The resulting large-$N$ phase diagrams are quite a bit more involved than the minimal scenarios reviewed above (and summarized in figures~\ref{BasicDiagram} and~\ref{TriangleD}): the theory develops~$N_f +1$ different vacuum superselection sectors, each with a different pattern of~$U(N_f)$ symmetry breaking. The resulting NG bosons are generally accompanied by a non-trivial Chern-Simons TQFT. These vacua are exactly degenerate at leading nontrivial order in the large-$N$ limit, and they are separated by an~$\CO(N)$ potential barrier. At subleading order the degeneracy is split, leading to a multitude of low-lying metastable vacua in large-$N$ QCD$_3$. Dialing the flavor-singlet mass parameter~$m$ of the theory leads to a sequence of first-order transitions that traverses each of these vacua in turn. The existence of these metastable vacua, as well as the resulting first-order phase transitions, is reminiscent of theta vacua in four-dimensional gauge theories, even though three-dimensional gauge theory does not admit a theta angle.\footnote{~The analogy between the results in this paper and four-dimensional  theta vacua can be made more precise by thinking about interfaces in four dimensions. Interfaces in four-dimensional large-$N$ QCD with light quarks exhibit a rich structure of phases and transitions, because the four-dimensional theory contains a light~$\eta'$ particle in the large-$N$ limit. There are many possible trajectories for the~$\eta'$ in field space as we cross the interface, and we expect these to describe the metastable three-dimensional vacua analyzed here. It would be nice to establish this in detail, by generalizing the analysis in the appendix of~\cite{Gaiotto:2017tne} (see also~\cite{Cherman:2017dwt,Argurio:2018uup}). This picture suggests that when~$N$ is sufficiently small, so that the~$\eta'$ is sufficiently heavy, the phase diagrams we consider in this paper may simplify.}  It is an attractive feature of the large-$N$ limit that we are able to explore both the regime~$k\geq {N_f\over 2 }$ and the regime~$0\leq k <{N_f \over 2}$ in a uniform manner, using a single set of tools.

Given that large-$N$ QCD$_3$ has multiple different phases, with NG bosons and TQFTs in the deep IR, which are connected by first-order phase transitions, it is interesting to ask what role the bosonic duals~\eqref{eq:introdual}, \eqref{TwoDuals} play in this context. Instead of providing an alternative description for a putative second-order transition, the bosonic duals capture all the IR phases of large-$N$ QCD$_3$ through different condensates of the scalar fields~$\phi$ and~$\hat \phi$, i.e.~different patterns of color-flavor locking. These condensates can be achieved by specifying a scalar potential with certain special properties, which we describe in detail. Interestingly, it is not possible to achieve these properties with a purely quartic scalar potential, which was previously assumed in most discussions of duality (see~\cite{Aharony:2018pjn} for a notable exception). The same scalar potential can also describe the first-order transitions between the different phases. 

It is tempting to speculate that duality has more to say about the physics of large-$N$ QCD$_3$ than simply matching the physics in the deep IR of its various phases. For instance, the presence of exactly degenerate vacua in the large-$N$ theory implies the existence of domain walls that interpolate between these vacua. Perhaps the duality can be used to infer the light degrees of freedom residing on the walls. Moreover, even though large-$N$ QCD$_3$ only displays first-order transitions, it is conceivable that it could be deformed in such a way as to give rise to a fixed point at complex couplings. In this case the duality could furnish a fully equivalent description of the resulting complex CFT (see for instance~\cite{Kaplan:2009kr, Gorbenko:2018ncu,Benvenuti:2018cwd} for related ideas). 

This paper is organized as follows: in section \ref{sec:largenqcd3} we analyze the vacua of QCD$_3$ at leading non-trivial order in the large-$N$ limit. In section \ref{sec:oneovern} we incorporate the leading~$1 \over N$ corrections, and in section \ref{sec:duality} we show that the phenomena uncovered in QCD$_3$ can be reproduced using its bosonic duals. 

{\it Comment Added:}\footnote{~We thank O.~Aharony for an illuminating exchange and ongoing discussions.}  Throughout this paper, we only analyze the limit of fixed $k$ and large $N$, but in fact our arguments extend to the regime of large~$k$ and large~$N$ with fixed, but large~$\lambda = {N \over k} \gg 1$. This superficially seems to contradict the literature on Chern-Simons-Matter theories, which flow to a CFT in this regime; by contrast, we predict a sequence of first-order transitions. This contradiction is resolved by recalling that we are not dropping the Yang-Mills kinetic term in our analysis. Intuitively, dropping the Yang-Mills term eliminates the gluon, which acquires a mass~$m_g \sim k g^2$ due to the Chern-Simons term. However, the full Yang-Mills-Chern-Simons theory becomes strongly coupled at the scale~$\Lambda = g^2N$. Thus we can only straightforwardly integrate out the gluon if~$m_g \gg \Lambda$, or equivalently if~$\lambda = {N \over k } \ll 1$.  In this small-$\lambda$ regime, the Yang-Mills-Chern-Simons theory reliably flows to the CFT uncovered in previous work (see above). Conversely, the dynamics of the Chern-Simons-Matter theories without a Yang-Mills term does not necessarily shed light on the behavior of QCD$_3$ with large~$\lambda$.

\section{The Large-$N$ Limit of QCD$_3$}\label{sec:largenqcd3}

In this section we analyze QCD$_3$, i.e.~$SU(N)_k$ Yang-Mills-Chern-Simons theory coupled to~$N_f$ flavors of fundamental quarks, at leading order in the large-$N$ expansion, while keeping~$k$ and~$N_f$ fixed. Throughout, we use two-component Dirac fermions. Gauge fields are hermitian~$A = A^a t^a$ with~$\Tr(t^a t^b) = \half \delta^{ab}$. We use Lorentzian signature for the metric.

\subsection{Yang-Mills-Chern-Simons Theory}\label{ssec:ymcs}

We begin by considering Yang-Mills theory with gauge group~$SU(N)$ and a level-$k$ Chern-Simons term, 
\begin{equation}\label{eq:ymcs}
\SL_\text{YMCS} = {1 \over 4g^2} \, \Tr\left(F \wedge \star F\right) + {k \over 4 \pi} \, \Tr \left( A \wedge dA + {2 \over 3} A \wedge A \wedge A\right)~, \qquad k \in \Z~.
\end{equation}
Here~$g^2$ is the three-dimensional Yang-Mills gauge coupling, which carries dimensions of mass. By contrast, $k \in \Z$ is the dimensionless, quantized Chern-Simons level. The theory~\eqref{eq:ymcs} has a~$\Z_N$ 1-form global symmetry associated with the center of~$SU(N)$ gauge group (see~\cite{Gaiotto:2014kfa,Kapustin:2014gua}). If~$k = 0$ it is also invariant under time-reversal symmetry~$\mathsf{T}$. 

In the absence of the Chern-Simons term (i.e.~when~$k = 0$), the theory in~\eqref{eq:ymcs} is expected to confine and flow to a gapped theory with a unique vacuum. In such a vacuum, all Wilson loop operators obey an area law, so that the~$\Z_N$ 1-form symmetry is unbroken. As in four dimensions~\cite{tHooft:1973alw}, the large-$N$ limit is defined by taking
\begin{equation}\label{eq:largenlimit}
N \; \rightarrow \; \infty~, \qquad  \Lambda = g^2 N = \text{fixed}~.
\end{equation}
Here the mass scale~$\Lambda$, which is the analogue of the four-dimensional 't Hooft coupling, furnishes the strong-coupling scale of the three-dimensional large-$N$ theory. Since~$g^2 \sim {1 \over N}$, it follows that gluon propagators scale as~$1 \over N$, while all vertices scale as~$N$. This leads to the usual perturbative large-$N$ counting rules, e.g.~planar vacuum diagrams scale as~$N^2$.   

Let us consider the effect of a non-zero Chern-Simons level~$k$ in~\eqref{eq:ymcs}. Since the Chern-Simons term has fewer derivatives than the Yang-Mills kinetic term, it is the leading term in the deep IR. In its presence, the gluons acquire a mass~$m_\text{gluon} \sim k g^2$. At low energies, the gauge degrees of freedom are described by pure~$SU(N)_k$ Chern-Simons theory, which is a TQFT. The line operators of this TQFT, which descend from the Wilson lines of the~$SU(N)$ Yang-Mills theory in the UV, have non-trivial correlation functions at long distances. Therefore the~$\Z_N$ 1-form symmetry is spontaneously broken and the gauge theory is in a deconfined phase. 

If we take the large-$N$ limit~\eqref{eq:largenlimit} while keeping~$k$ fixed, then the effects of the Chern-Simons term are formally subleading. Intuitively, this follows from the fact that the gluon mass~$m_\text{gluon} \sim k g^2 \sim {k \Lambda \over N}$ vanishes at leading order in the large-$N$ expansion. Alternatively, we can rescale the~$SU(N)$ gauge field~$A \rightarrow {1 \over \sqrt N} B$, so that the Lagrangian~\eqref{eq:ymcs} takes the following schematic form,
\begin{equation}\label{eq:rescymcs}
\SL \sim {1 \over \Lambda} \Tr \left( \left(dB\right)^2 + {1 \over \sqrt N} \,  B^2 dB + {1 \over N} \, B^4\right) + {k \over N} \Tr \left(B dB\right)  + {k \over N^{3/2}} \Tr \left( B^3\right)~.
\end{equation}
For the purpose of counting factors of~$N$, replacing a Yang-Mills propagator or three-point vertex by their Chern-Simons counterparts therefore multiplies a given diagram by~$1 \over N$. 

The fact that the Chern-Simons term is subleading in the large-$N$ expansion does not mean that it can be completely ignored. For instance, when~$k \neq 0$ it is not correct to conclude that the large-$N$ theory leads to a confining, gapped theory with a unique, trivial vacuum, as is the case for the~$k = 0$ theory. Instead, as reviewed above, the theory flows to a deconfined~$SU(N)_k$ Chern-Simons TQFT in the deep IR, below the scale~${k \Lambda \over N}$ set by the gluon mass, even though this scale vanishes in the large-$N$ limit. This is similar to large-$N$ QCD in four dimensions with a finite number of massless quarks: the massless NG bosons that appear in the deep IR arise from the quarks, which only enter at the first subleading order in~$1 \over N$. We will encounter an analogous phenomenon in section~\ref{ssec:vacefts}.\footnote{~Another subtle issue is that certain observables in Chern-Simons theory that are sufficiently sensitive to global issues can violate the standard perturbative large-$N$ counting rules. For instance, the free energies on compact Euclidean manifolds such as~$S^3$ and~$T^3$ are  enhanced by factors of order~$\log N$. (The partition function on a large torus counts the number of ground states. For instance, if $k=1$, this is just $N$, and hence the free energy is proportional to $\log N$.) Similar violations of the naive large-$N$ counting rules occur on other manifolds~\cite{Witten:1988hf}. Intuitively, this is possible because the Chern-Simons term is not a gauge-invariant local operator. This peculiarity will not affect our discussion.}

\subsection{Adding Flavors}\label{ssec:flavors}

We now add~$N_f$ flavors of quarks, i.e.~complex, two-component Dirac fermions~$\psi^i$ (with~$i = 1, \ldots, N_f$) that transform in the fundamental representation of the~$SU(N)$ gauge symmetry, to the Yang-Mills-Chern-Simons action~\eqref{eq:ymcs},
\begin{equation}\label{eq:qcd3lag}
\SL = {1 \over 4 g^2} \, \Tr\left(F \wedge \star F\right) + {k \over 4 \pi} \, \Tr \left( A \wedge dA + {2 \over 3} A \wedge A \wedge A\right)+ i \b \psi_i \slashed{D}_A \psi^i - {m_i}^j \, \b \psi_j \psi^i~.
\end{equation}
Here~$D_A = d - i A$ is the gauge-covariant derivative, while the~$\psi^i$ are Dirac spinors in three dimensions. The mass parameters~${m_i}^j$ satisfy the reality condition~$\left({m_i}^j\right)^* = {m_j}^i$, so that the~$N_f \times N_f$ mass matrix~$\mathbf m$ (whose~$i$-$j$ entry is given by~${m_i}^j$) is hermitian,
\begin{equation}\label{eq:mmatherm}
{\mathbf m}^\dagger = {\mathbf m}~.
\end{equation}
The presence of the~$N_f$ fundamental fermions modifies the quantization condition for the Chern-Simons level~$k$ in~\eqref{eq:ymcs} as follows,
\begin{equation}\label{eq:csquant}
k + \half \, N_f \in \Z~.
\end{equation}

Let us discuss the global symmetries of the~QCD$_3$ Lagrangian~\eqref{eq:qcd3lag}: 
\begin{itemize}
\item When the mass matrix~$\mathbf m$ vanishes, there is a~$U(N_f)$ flavor symmetry under which the~$\psi^i$ transform in the fundamental representation. The central~$U(1)_B \subset U(N_f)$ is the baryon number symmetry. It is preserved by any mass matrix~$\mathbf m$, while the~$SU(N_f) \subset U(N_f)$ flavor symmetry is generically broken. Since~$\mathbf m$ is a hermitian matrix (see~\eqref{eq:mmatherm}), we can use an~$SU(N_f)$ transformations to diagonalize it, with real eigenvalues~$m_i$, 
\begin{equation}\label{eq:mmatdiag}
\mathbf m =  \text{diag}\left(m_1, \ldots, m_{N_f}\right)~, \qquad m_i \in \R~.
\end{equation}
Note that the~$m_i$ can have either sign. This is typical in three dimensions and will play an important role below. Note that we can use an~$SU(N_f)$ Weyl transformation to arrange the~$m_i$ in descending order,
\begin{equation}\label{eq:meigorder}
m_1 \geq m_2 \geq \cdots \geq  m_{N_f}~.
\end{equation}
If all eigenvalues~$m_i$ are distinct, there is only an unbroken~$U(1)^{N_f}$ symmetry. By contrast, if all~$m_i$ coincide the mass matrix is given by
\begin{equation}\label{eq:flavsingm}
\mathbf m = m \, \1~, \qquad m \in \R~,
\end{equation}
where~$\1$ is the~$N_f \times N_f$ identity matrix.  Such a mass term preserves the full~$U(N_f)$ flavor symmetry. For this reason, we will refer to it as a flavor-singlet mass.

\item We can define a time-reversal operation~$\mathsf T$ that commutes with the~$U(N_f)$ flavor symmetry, so that both the Chern-Simons level~$k$ and the mass matrix~$\mathbf m$ (and hence its entries~${m_i}^j$, as well as its eigenvalues~$m_i$ in~\eqref{eq:mmatdiag}) are odd under~$\mathsf T$. Therefore the theory is~$\mathsf T$-invariant if~$k$ and~$\mathbf m$ vanish. Note that certain non-vanishing configurations of $\mathbf m$ are preserved by a combination of~$\mathsf T$ and an element of $U(N_f)$ (see footnote~\ref{lab:TF} for an example). 
\end{itemize}
Note that the~$\Z_N$ 1-form global symmetry that was present in the Yang-Mills-Chern-Simons theory~\eqref{eq:ymcs} is explicitly broken by the inclusion of the fundamental quarks, which serve as endpoints for Wilson lines in the fundamental representation of~$SU(N)$. 

The quark mass matrix~$\mathbf m$ decomposes into a traceless part~$\hat {\mathbf m}$, as well as the trace part~$m \, \1$ in~\eqref{eq:flavsingm}. Since~$\hat {\mathbf m}$ transforms in the adjoint representation of~$SU(N_f) \subset U(N_f)$, the point~$\hat {\mathbf m} = 0$ is the distinguished point in parameter space where the full~$SU(N_f)$ symmetry is restored. The flavor-singlet mass~$\mathbf m= m \, \1$ in~\eqref{eq:flavsingm} preserves the entire~$U(N_f)$ flavor symmetry, as we said. As long as~$k = 0$, the massless point~$m = 0$ is the distinguished point in parameter space where time-reversal symmetry~$\mathsf T$ (as well as the full flavor symmetry) is restored. This ceases to hold when~$k \neq 0$ and~$\mathsf T$ is no longer a global symmetry for any value of~$m$. In this case the flavor-singlet mass~$m$ is additively renormalized by the Yang-Mills-Chern-Simons interactions, and specifying the point in parameter space where~$m = 0$ amounts to a choice of regularization scheme.\footnote{~This is reminiscent of what happens in four-dimensional QCD with~$N_f = 1$ fundamental Dirac flavor, whose mass term also does not break any global symmetries of the massless theory. See for instance the recent discussion in~\cite{Gaiotto:2017tne}.} 

We can say more in the large-$N$ limit with fixed~$k$ and~$N_f$ (see section~\ref{ssec:ymcs}). Since the breaking of time-reversal~$\mathsf T$ by the Chern-Simons term~$k$ is subleading in the large-$N$ expansion, the additive renormalization~$\Delta m$ of~$m$ due to this term scales as~$\Delta m \sim {k \Lambda \over N}$. The location of the origin~$m = 0$ is therefore well defined to within this accuracy. 

\subsection{The Disk Diagram: Degenerate Vacua and Symmetry Breaking}\label{ssec:disksb}

Consider the gauge-invariant, hermitian meson operator
\begin{equation}\label{eq:mesondef}
{M_i}^j = {1 \over N} \, \b \psi_i \psi^j~, \qquad \left({M_i}^j\right)^\dagger = {M_j}^i~.
\end{equation}
Its transformation properties under the~$U(N_f)$ flavor symmetry and time-reversal~$\mathsf T$ are the same as those of the mass parameters~${m_i}^j$ in~\eqref{eq:qcd3lag}, i.e.~it transforms in the adjoint representation of~$SU(N_f) \subset U(N_f)$, it is invariant under the central~$U(1)_B \subset U(N_f)$ baryon number symmetry, and it is~$\mathsf T$-odd. As was the case there, it is convenient to introduce an~$N_f \times N_f$ hermitian matrix~$\mathbf M = {\mathbf M}^\dagger$, whose entries are given by~${M_i}^j$. Whenever (a subgroup of)~$U(N_f)$ or~$\mathsf T$ are global symmetries of the QCD$_3$ Lagrangian~\eqref{eq:qcd3lag}, the meson vacuum expectation value (vev)~$\langle \mathbf M \rangle$ is therefore an interesting order parameter for their spontaneous breaking. We begin by analyzing the massless theory, with~$\mathbf m = 0$, whose Lagrangian preserves the entire~$U(N_f)$ flavor symmetry. When~$k = 0$, it also preserves time reversal~$\mathsf T$. 

We will follow the four-dimensional discussion of~\cite{Coleman:1980mx}, as well as the work~\cite{Ferretti:1992fd,Ferretti:1992fga,Hong:2010sb} in three dimensions, to construct the effective potential~$V\big(\langle \mathbf M\rangle\big)$ for the meson vev~$\langle \mathbf M\rangle$ at low orders in the large-$N$ expansion. As in these papers, we assume that~$\langle \mathbf M\rangle$ is the correct order parameter to diagnose the spontaneous breaking of the global~$U(N_f)$ or~$\mathsf T$ symmetries. Due to the transformation properties of the meson~$\mathbf M$ under these symmetries, its vev~$\langle \mathbf M\rangle$ can only account for certain symmetry-breaking patterns. In particular, $\mathbf M$ is invariant under the~$U(1)_B$ baryon number symmetry, and hence this symmetry will necessarily always be unbroken in our analysis. It is possible to contemplate more general order parameters and breaking patterns,\footnote{~For instance, four-dimensional~$SU(2)$ gauge theory with two Weyl fermions~$\lambda^{i = 1,2}_\alpha$ in the adjoint representation of the gauge group has a continuous~$SU(2)$ flavor symmetry acting on the doublet index~$i$, and a discrete~$\Z_8$ flavor symmetry which acts on~$\lambda^i$ by eighth roots of unity. Standard chiral symmetry breaking by a fermion bilinear~$\lambda^{(i} \lambda^{j)}$ spontaneously breaks~$SU(2) \rightarrow U(1)$ and~$\Z_8 \rightarrow \Z_4$, see for instance~\cite{Cordova:2018acb} and references therein. A more exotic possibility discussed in~\cite{Anber:2018tcj,Cordova:2018acb} is that the~$SU(2)$ flavor symmetry remains unbroken, but the~$\Z_8 \rightarrow \Z_4$ breaking persists. This breaking pattern can be diagnosed using a four-fermion order parameter~$\lambda^{(i} \lambda^{j)} \lambda_{(i} \lambda_{j)}$. Scenarios with unbroken~$\Z_8$ symmetry are discussed in~\cite{Cordova:2018acb, Bi:2018xvr}.
Symmetry breaking by four-fermion operators in four-dimensional QCD was discussed recently in~\cite{Tanizaki:2018wtg}.} but we will not do so here. 

The construction of the effective potential~$V\big(\langle \mathbf M\rangle\big)$ is standard. First we turn on generic mass parameters~${m_i}^j$ in~\eqref{eq:qcd3lag}, which act as sources for the meson  operators~\eqref{eq:mesondef},
\begin{equation}\label{eq:massterm}
\SL \quad \supset \quad -{m_i}^j \b \psi_j \psi^i = -N {m_i}^j {M_j}^i = -N \tr \left(\mathbf m \,  \mathbf M\right)~.
\end{equation}
Here~$\tr$ denotes a trace over the flavor indices~$i, j = 1, \ldots, N_f$. The mass term~\eqref{eq:massterm} completely breaks any vacuum degeneracy that can be detected by the order parameter~$\langle \mathbf M\rangle$. In line with the discussion between~\eqref{eq:mesondef} and~\eqref{eq:massterm} above, we will assume that this is the only source of vacuum degeneracy. 

We now compute the partition function~$Z(\mathbf{m})$ as a function of the mass matrix~$\mathbf m$. In Lorentzian signature, we can express~$Z(\mathbf m) = \exp\left(i \, V_3 W(\mathbf m)\right)$, where~$V_3$ is the spacetime volume (which appears because the sources in~\eqref{eq:massterm} are translationally invariant) and the function~$W(\mathbf m)$ only receives contributions from connected correlation functions of the meson operator~$\mathbf M$. Note that~$W(\mathbf m)$ has mass-dimension three. It follows from~\eqref{eq:massterm} that
\begin{equation}\label{eq:mvev}
N \langle \mathbf M\rangle = -{\d W(\mathbf m) \over \d \mathbf m}~.
\end{equation}

\begin{figure}[!th]
\centerline{\includegraphics[scale=0.3]{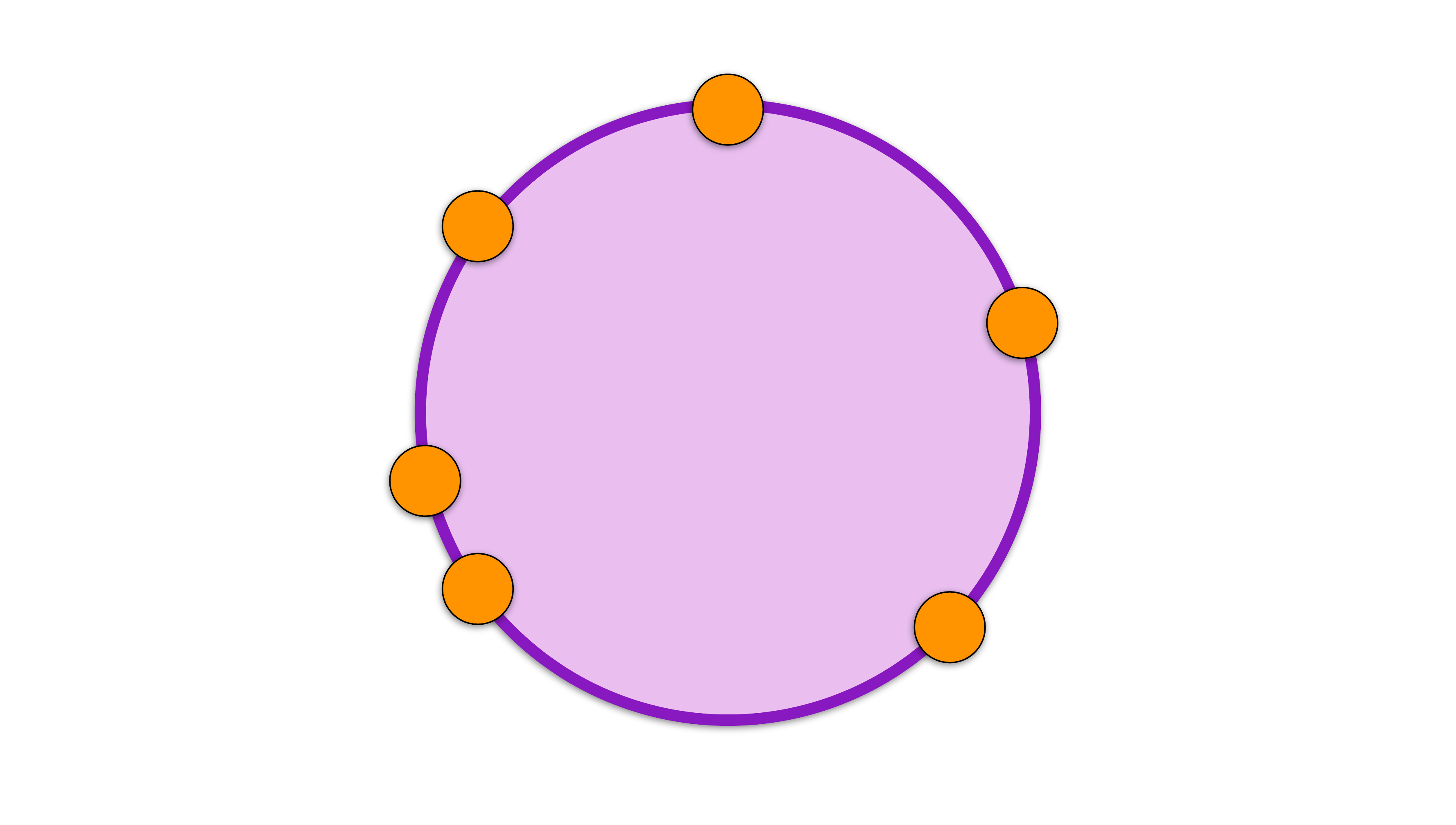}}
\caption{Disk diagram. The orange circles on the boundary represent the insertion of meson operators.}
\label{disk}
\end{figure}

The diagrams that contribute to connected meson correlators (and hence~$W(\mathbf m)$) at the lowest non-trivial order in the large-$N$ expansion have the topology of a disk. Each such disk diagram has a single fermion loop at the boundary, where all meson operators~$\mathbf M$ are inserted, while the interior is filled in by arbitrary planar gluon diagrams, see figure~\ref{disk}.

These disk diagrams have several important properties:
\begin{itemize}
\item[1.)] If all gluon vertices and propagators are of Yang-Mills type, the disk diagrams scale like~$N$ in the large-$N$ limit. As discussed in section~\ref{ssec:ymcs}, the insertion of any Chern-Simons vertices or propagators leads to diagrams that are subleading in the large-$N$ expansion. At leading order we can therefore ignore the Chern-Simons level~$k$.
 
\item[2.)] Since disk diagrams only involve a single fermion loop, they can only lead to single-trace terms in~$W(\mathbf m)$. The leading~$\CO(N)$ part of~$W(\mathbf m)$ therefore schematically takes the following form,
\begin{equation}\label{eq:singletr}
W(\mathbf m)\big|_{\CO(N)} = N \Lambda^3 \sum_{n = 1}^\infty {c_n \over \Lambda^n} \, \tr\left({\mathbf m}^n\right)~.
\end{equation}
Here~$\Lambda$ is the strong-coupling scale of the large-$N$ theory, which was defined in~\eqref{eq:largenlimit}, and~$n$ runs over the number of meson operators~$\mathbf M$ inserted at the boundary of the disk diagram. Since the gauge interactions are flavor blind and we can effectively set~$k = 0$ at the level of the~$\CO(N)$ disk diagrams (see point~$1.)$ above), the dimensionless coefficients~$c_n$ do not depend on~$N$, $N_f$ or~$k$. The series expansion on the right-hand side of~\eqref{eq:singletr} should not be taken literally, because~$W({\bf m})$ is not analytic around the massless point. Rather, the main point of~\eqref{eq:singletr} is to emphasize the single-trace nature of the diagrams contributing to~$W({\bf m})$.

\item[3.)] Since the~$\CO(N)$ disk diagrams are insensitive to $k$ (see point~$1.)$ above), they are constrained by time-reversal symmetry~$\mathsf T$ (see the discussion below~\eqref{eq:flavsingm}). Note that this statement applies for all values of~$N_f$, even though the quantization condition~\eqref{eq:csquant} forbids theories with vanishing~$k$ and odd~$N_f$. Since the mass and meson matrices~$\mathbf m$ and~$\mathbf M$ are~$\mathsf T$-odd, we conclude that~$W(\mathbf m)\big|_{\CO(N)}$ must be an even function of~$\mathbf m$,
\begin{equation}\label{eq:evenwofm}
W(- \mathbf m)\big|_{\CO(N)} = W(\mathbf m)\big|_{\CO(N)}~.
\end{equation}
Therefore the sum over meson insertions in~\eqref{eq:singletr} can be restricted to even positive integers. More precisely, only even, single trace functions of the eigenvalues are allowed.

\end{itemize}

\noindent Given~$W(\mathbf m)$, the effective potential~$V\big(\langle \mathbf M\rangle \big)$ is defined by a Legendre transform,
\begin{equation}\label{eq:effvdef}
V\big(\langle \mathbf M\rangle \big) = - W(\mathbf m) - N  \tr\left(\mathbf m \, \mathbf M\right)~.
\end{equation}
It follows from~\eqref{eq:mvev} that this effective potential is independent of~$\mathbf m$, but that it depends on the meson vev~$\langle \mathbf M\rangle$ once we express~$\mathbf m$ in terms of~$\langle \mathbf M\rangle$. At~$\CO(N)$ in the large-$N$ expansion, $W(\mathbf m)$ is equal to a sum~\eqref{eq:singletr} of single-trace terms~$\sim N \tr\left({\mathbf m}^n\right)$ with even~$n$. Together with~\eqref{eq:mvev}, this implies that the meson vev~$\langle {\mathbf M} \rangle$ is~$\CO(1)$ in the large-$N$ limit. It also implies that the~$\CO(N)$ contribution to the effective potential~\eqref{eq:effvdef} can be written as a sum over even single-trace terms involving the meson vev~$\langle \mathbf M\rangle$,
\begin{equation}\label{eq:largenveff}
V\left(\langle \mathbf M\rangle \right)\big|_{\CO(N)} = N \Lambda^3 \sum_{n'  = 2, 4, 6, \ldots}^\infty {c'_n \over \Lambda^{2n}} \, \tr\left(\langle \mathbf M\rangle^n\right)~.
\end{equation}
The dimensionless coefficients~$c'_n$ do not depend on~$N$, $N_f$ or~$k$. They can be determined from the analogous coefficients~$c_n$ in~\eqref{eq:singletr} via the Legendre transform~\eqref{eq:effvdef}. 

In order to analyze the effective potential~\eqref{eq:largenveff}, it is convenient to use an~$SU(N_f)$ transformation to diagonalize the hermitian meson vev,
\begin{equation}\label{eq:mesmatdiag}
\langle \mathbf M\rangle = \Lambda^2 \, \text{diag} \left(x_1, \ldots, x_{N_f}\right)~, \qquad x_i \in \R~, \qquad i = 1, \cdots, N_f~.
\end{equation}
Here the~$x_i$ are the dimensionless eigenvalues of~$\langle \mathbf M\rangle$. Since both~$\langle \mathbf M\rangle$ and~$\Lambda$ are~$\CO(1)$ in the large-$N$ limit, the same is true of the~$x_i$. As was the case for the eigenvalues~$m_i$ of the mass matrix in~\eqref{eq:mmatdiag}, it is crucial that the meson eigenvalues~$x_i$ can have either sign. Substituting into~\eqref{eq:largenveff}, we find that the large-$N$ effective potential has the form
\begin{equation}\label{eq:veffviafx}
V(x_i) \big|_{\CO(N)} = N \Lambda^3 \sum_{i = 1}^{N_f} F(x_i)~.
\end{equation}
Here~$F(x)$ is the following even function of~$x$, 
\begin{equation}\label{eq:fxeven}
F(x) = \sum_{n' = 2, 4, 6, \ldots}^\infty c'_n \, x^n = F(-x)~.
\end{equation}
Note that the different eigenvalues~$x_i$ in~\eqref{eq:veffviafx} are all independent and do not interact with each other. This is an immediate consequence of the single-trace structure of the effective potential~\eqref{eq:largenveff}. 

\begin{figure}[!th]
\centerline{\includegraphics[scale=1]{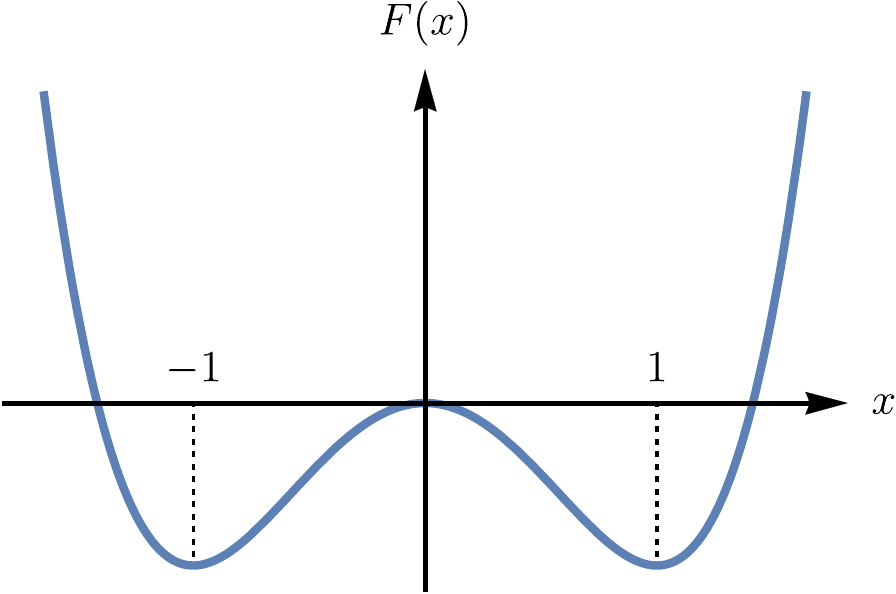}}
\caption{The main assumption of this paper is that the function~$F(x)$ that determines the~$\CO(N)$ effective potential for the eigenvalues of the meson one-point function~$\langle \mathbf M\rangle$ is minimized away from the origin, and hence takes the form of a symmetry-breaking Ising-like potential. We rescale~$x$ so that the minima of~$F(x)$ are located at~$x = \pm 1$.}
\label{fig:GraphIsing}
\end{figure}

In the absence of accidental degeneracies that are unrelated to spontaneous symmetry breaking, there are only two possibilities for the minima of the function~$F(x)$ in~\eqref{eq:fxeven}:
\begin{itemize}
\item[a)] $F(x)$ has a unique minimum at~$x = 0$. In this case the effective potential~\eqref{eq:veffviafx} is minimized at~$\langle \mathbf M\rangle = 0$ and there is no spontaneous symmetry breaking. A plausible scenario is that in this case the theory flows to a CFT in the deep IR.
  
\item[b)] $F(x)$ has a pair of degenerate minima away from the origin. It therefore takes the form of an Ising-like potential in the symmetry-breaking phase, as depicted in figure~\ref{fig:GraphIsing}. In this case the effective potential~\eqref{eq:veffviafx} is necessarily minimized at~$\langle \mathbf M \rangle \neq 0$, which leads to spontaneous symmetry breaking in QCD$_3$. By rescaling the meson~$\mathbf M$, and hence its eigenvalues~$x_i$ in~\eqref{eq:mesmatdiag}, we are free to place the minima of~$F(x)$ at~$x = \pm1$. 
\end{itemize}

\noindent There is a reasonable amount of evidence in favor of option~b) above, see for instance~\cite{Karthik:2016bmf,Kanazawa:2019oxu} and references therein. For the remainder of this paper, we will assume that option~b)~is indeed realized. All of our conclusions below will follow from (and therefore hinge on) this key assumption.  

It is now straightforward to determine the vacuum structure at leading order in the large-$N$ expansion. This can be done in two steps:
\begin{itemize}
\item[1.)] Find the number of inequivalent ways to minimize the~$\CO(N)$ effective potential~\eqref{eq:veffviafx}. Here the equivalence classes consist of configurations of the~$x_i$ that are related by an~$SU(N_f)$ transformation. 
\item[2.)] Within each equivalence class, determine the pattern of spontaneous~$SU(N_f)$ breaking. The resulting manifold of NG bosons determines the vacua within each equivalence class, which are related by the spontaneously broken part of the~$SU(N_f)$ symmetry.   
\end{itemize}
In order to carry out step~1.), it is convenient to use the remaining~$SU(N_f)$ Weyl transformations to order the meson eigenvalues~$x_i$ in ascending order, 
\begin{equation}\label{eq:meseigorder}
x_1 \leq x_2 \leq \cdots \leq x_{N_f}~.
\end{equation}
Since the function~$F(x)$ that determines the~$\CO(N)$ effective potential~\eqref{eq:veffviafx} is minimized at~$x = \pm 1$, every meson eigenvalue~$x_i$ can reside at either minimum. Taking into account the ordering~\eqref{eq:meseigorder}, we therefore find~$N_f + 1$ equivalence classes of vacua. We choose to label these equivalence classes by the number~$p \in \{0, 1, \ldots, N_f\}$ of eigenvalues that reside in the~$x = -1$ vacuum, 
\begin{equation}\label{eq:pvacdef}
x_1 = \cdots = x_p = -1~, \qquad x_{p+1} = \cdots = x_{N_f} = 1~.
\end{equation}

We can now carry out step~2.) described above. 
The eigenvalue configuration~\eqref{eq:pvacdef} spontaneously breaks the~$U(N_f)$ flavor symmetry as follows,
\begin{equation}\label{eq:flavsymbreak}
U(N_f) \quad \rightarrow \quad U(p) \times U(N_f - p)~, \qquad p \in \{0, 1, \ldots, N_f\}~.
\end{equation}
As was already explained above~\eqref{eq:massterm}, the central~$U(1)_B \subset U(N_f)$ baryon number symmetry necessarily remains unbroken. The NG manifold resulting from~\eqref{eq:flavsymbreak}, which determines the vacua in the equivalence class labeled by~$p$, is therefore a complex Grassmannian,
\begin{equation}\label{eq:grassdef}
\text{Gr}(p, N_f) = {U(N_f) \over U(p) \times U(N_f - p)}~, \qquad p \in \{0, 1, \ldots, N_f\}~.
\end{equation}

Let us summarize the picture for the vacuum structure that we have assembled so far, which holds at leading~$\CO(N)$ order in the large-$N$ expansion. As we have emphasized above, this picture does not depend on the Chern-Simons level~$k$:

\begin{itemize}
\item There are~$N_f+1$ degenerate superselection sectors, labeled by an integer\\ $p \in \{0, 1, \ldots, N_f\}$, which are not related by the~$U(N_f)$ flavor symmetry. 
\item Within each superselection sector, the~$U(N_f)$ flavor symmetry is spontaneously broken as in~\eqref{eq:flavsymbreak}, leading to the Grassmannian NG manifold~$\text{Gr}(p, N_f)$ in~\eqref{eq:grassdef}. 

\item Since the leading-order effective potential scales like~$N$, it follows that vacua in two distinct superselection sectors are separated by an energy barrier of order~$N$. This simple observation will have important consequences below.
\end{itemize}
\noindent The possibility of degenerate superselection sectors in the large-$N$ limit was also noted in~\cite{Ferretti:1992fd,Ferretti:1992fga,Hong:2010sb,Kanazawa:2019oxu}.

We would like to comment on the behavior of the superselection sectors described above under the action of time reversal~$\mathsf T$, which is a symmetry of the massless QCD$_3$ Lagrangian~\eqref{eq:qcd3lag} when~$k = 0$ and~$N_f$ is even (see~\eqref{eq:csquant}). Since the meson~$\mathbf M$, and hence its eigenvalues~$x_i$, are~$\mathsf T$-odd, it follows from~\eqref{eq:pvacdef} that time reversal exchanges the superselection sectors~$p \; \leftrightarrow \; N_f - p$. There are therefore~$N_f \over 2$ pairs of superselection sectors that are exchanged by the spontaneously-broken~$\mathsf T$-symmetry. The only exception is the superselection sector labeled by~$p = {N_f \over 2}$, which maps to itself under time-reversal symmetry. The NG manifold in this sector is given by~$\text{Gr}\left({N_f \over 2} , N_f \right) = {U(N_f) \over U\left({N_f \over 2}\right) \times U\left({N_f \over 2}\right) }$. The two~$U\left({N_f \over 2}\right)$ factors, which correspond to the~$N_f \over 2$ negative and the~$N_f \over 2$ positive eigenvalues of~$\langle \mathbf M \rangle$, are exchanged by~$\mathsf T$. This exchange, which ruins the ordering of the eigenvalues in~\eqref{eq:meseigorder}, can be undone by a Weyl transformation~$\mathsf \CF \in SU(N_f)$. Therefore both~$\mathsf T$ and~$\CF$ are spontaneously broken in the~$p = {N_f \over 2}$ sector, but their product~$\mathsf T \CF$ is preserved. Note that the existence of the superselection sectors with~$p \neq {N_f \over 2}$ is superficially in tension with the work of~\cite{Vafa:1983tf,Vafa:1984xh}, where the authors argued for the symmetry-breaking pattern~$U(N_f) \;  \rightarrow \; U\left({N_f \over 2}\right) \times U\left({N_f \over 2} \right)$. This will be further discussed in sections~\ref{ssec:massdeform} and~\ref{ssec:annulus} below.

Another important aside concerns anomaly matching. The massless $k=0$ theory has a nonabelian parity anomaly, i.e.~a mixed~$\mathsf T$-$SU(N_f)$ 't Hooft anomaly, as long as~$N$ is odd~\cite{Niemi:1983rq,Redlich:1983dv}. It has been argued that this anomaly can only be matched by gapless degrees of freedom, not a TQFT~\cite{Wang:2014lca,Cordova:2019bsd,COprog}. This  is nicely consistent with the fact that our $\mathsf T$-invariant phase has massless NG bosons. 

\subsection{Mass Deformations}\label{ssec:massdeform}

Let us turn on quark masses~${m_i}^j$ in the QCD$_3$ Lagrangian~\eqref{eq:qcd3lag}. As long as these masses are small compared to the strong-coupling scale~$\Lambda$, we can analyze their effect by minimizing the leading-order contribution of the mass term~\eqref{eq:massterm} to the effective potential,
\begin{equation}\label{eq:smallmdef}
\Delta V\big(\langle \mathbf M\rangle\big) = N \tr \left( \mathbf m \, \langle \mathbf M\rangle\right)~,
\end{equation}
over the minima of the effective potential~$V\left(\langle \mathbf M\rangle\right)$ of the massless theory in~\eqref{eq:largenveff}. As discussed around~\eqref{eq:pvacdef}, these minima are described by hermitian matrices~$\langle \mathbf M\rangle$ whose eigenvalues are~$\pm1$, or equivalently hermitian matrices~$\langle \mathbf M\rangle$ that satisfy~$\langle \mathbf M\rangle^2 = \1$. 

In order to implement the constraint~$\langle \mathbf M\rangle^2 = \1$, we use a Lagrange multiplier~$\mathbf L$, which is itself a hermitian~$N_f \times N_f$ matrix. Our task is then to minimize the following function,
\begin{equation}\label{eq:lagmmulmin}
\tr \Big( \mathbf m \, \langle \mathbf M\rangle + \mathbf L \left( \langle \mathbf M\rangle^2 - \1\right)\Big)~.
\end{equation}
Extremizing with respect to~$\langle \mathbf M\rangle$ leads to 
\begin{equation}
\mathbf m + \langle \mathbf M\rangle \, \mathbf L + \mathbf L \, \langle \mathbf M\rangle = 0~.
\end{equation}
If we multiply this equation by~$\langle \mathbf M\rangle$ from the left and the right, subtract the two resulting equations,  and simplify using~$\langle \mathbf M\rangle^2 = \1$, we conclude that
\begin{equation}\label{eq:vacalign}
\left[\mathbf m \, , \, \langle \mathbf M\rangle\right] = 0~.
\end{equation}
This vacuum alignment condition states that meson vev~$\langle \mathbf M\rangle$ and the mass perturbation~$\mathbf m$ can be simultaneously diagonalized using a single~$SU(N_f)$ transformation. We will therefore diagonalize~$\mathbf m$ as in~\eqref{eq:mmatdiag} and~$\langle \mathbf M \rangle$ as in~\eqref{eq:mesmatdiag}. Substituting the eigenvalues~$m_i, x_i$ into~\eqref{eq:smallmdef}, we find that
\begin{equation}\label{eq:delveigenv}
\Delta V = N \Lambda^2 \sum_{i = 1}^{N_f} m_i x_i~.
\end{equation}

We must minimize~\eqref{eq:delveigenv} over all possible configurations~$x_i = \pm 1$ of the meson eigenvalues. If some of the~$m_i$ vanish, the corresponding~$x_i$ can take either value. As above, this leads to multiple degenerate vacua. Let us instead assume that all~$m_i \neq 0$, so that the degeneracy between the $N_f+1$ superselection sectors is lifted. As in~\eqref{eq:meigorder}, we can order the~$m_i$ in descending order and define an integer~$p$ to be the number of positive mass eigenvalues,
\begin{equation}\label{eq:pposms}
m_1 \geq m_2 \geq \cdots m_p > 0 > m_{p+1} \geq \ldots m_{N_f}~, \qquad p \in \{0, 1, \ldots, N_f\}~. 
\end{equation}
The mass deformation~\eqref{eq:delveigenv} is then minimized by the configuration of meson eigenvalues in~\eqref{eq:pvacdef}, which we repeat here,
\begin{equation}
x_1 = \cdots = x_p = -1~, \qquad x_{p+1} = \cdots x_{N_f} = 1~.
\end{equation}
The configuration of mass eigenvalues in~\eqref{eq:pposms} therefore singles out precisely one superselection sector, labeled by~$p$, from among the~$N_f + 1$ degenerate sectors described around~\eqref{eq:flavsymbreak} and~\eqref{eq:grassdef}. Moreover the mass deformation selects a unique vacuum on the Grassmannian target space $\text{Gr}(p, N_f)$.

If we dial any mass parameter~$m_i$ from negative to positive values, the corresponding meson eigenvalue~$x_i$ jumps from~$x_i = +1$ to~$x_i = -1$. At the massless point~$m_i = 0$, the two vacua are exactly degenerate. These transitions are therefore first order. Recall from the discussion at the end of section~\ref{ssec:flavors} that the point~$m_i = 0$ is well defined for all mass eigenvalues, and in particular for their sum~$\sum_i m_i$, as long as~$k = 0$ (so that the theory is invariant under time reversal~$\mathsf T$) or as long as we are working to leading non-trivial order in the large-$N$ expansion.

We would like to clarify the relation of the results obtained above, which hold at leading order in the large-$N$ limit, to the work of~\cite{Vafa:1983tf,Vafa:1984xh}. These authors considered massless~QCD$_3$ with~$k = 0$ and even~$N_f$, and they argued in favor of the symmetry breaking pattern~$U(N_f) \; \rightarrow \; U\left({N_f \over2}\right) \times U\left({N_f \over2}\right)$. In particular, they argued that the~$U\left({N_f \over2}\right) \times U\left({N_f \over2}\right)$ subgroup of the flavor symmetry is necessarily unbroken.  Among the~$N_f + 1$ degenerate superselection sectors we described around~\eqref{eq:flavsymbreak} and~\eqref{eq:grassdef}, only the superselection sector labeled by~$p = {N_f \over 2}$ has his property. 

The two pictures are reconciled by recalling that the arguments in~\cite{Vafa:1983tf,Vafa:1984xh} involve turning on time-reversal invariant fermion masses,\footnote{\label{lab:TF}~The time-reversal operation preserving these masses is not~$\mathsf T$, but rather what we called~$\mathsf T \CF$ at the end of section~\ref{ssec:disksb}.}
\begin{equation}\label{eq:vwmasses}
m_1 = \cdots = m_{N_f \over 2} = m ~, \qquad m_{{N_f \over 2} + 1} = \cdots = m_{N_f} = -m~,
\end{equation}
and taking the limit~$m \rightarrow 0^+$. The authors of~\cite{Vafa:1983tf,Vafa:1984xh} showed that the~$U\left({N_f \over2}\right) \times U\left({N_f \over2}\right)$ flavor symmetry preserved by the mass parameters~\eqref{eq:vwmasses} must remain unbroken in the massless limit. This is entirely consistent with our discussion above: the mass parameters in~\eqref{eq:vwmasses} correspond to~$p = {N_f \over 2}$ in~\eqref{eq:pposms}. In the massless limit, we therefore recover the superselection sector labeled by this value of~$p$, which exhibits the symmetry-breaking pattern derived in~\cite{Vafa:1983tf,Vafa:1984xh}. 

However, the existence of a unique~$U\left({N_f \over2}\right) \times U\left({N_f \over2}\right)$-preserving vacuum at non-vanishing~$m$ that can be tracked to~$m = 0$ does not exclude the possibility of additional vacua, with potentially different symmetry-breaking patterns, in the massless theory. For small non-zero~$m$, the additional vacua of the massless theory become low-lying metastable states. The possibility of accidental vacuum degeneracies, beyond the degeneracies expected from a particular pattern of spontaneous symmetry breaking, was noted in~\cite{Vafa:1983tf,Vafa:1984xh}, but deemed unlikely. Here we see that this scenario is in fact realized in~QCD$_3$, if we work at leading order in the large-$N$ expansion. In section~\ref{ssec:annulus} we will argue that these accidental degeneracies are lifted at the next-to-leading order, in accordance with the general expectations laid out in~\cite{Vafa:1983tf,Vafa:1984xh}.

\subsection{The Low-Energy Theories in the Different Vacua}\label{ssec:vacefts}

In section~\eqref{ssec:disksb} we analyzed the effective potential for the meson operator~$\mathbf M$ in massless~QCD$_3$ at leading order in the large-$N$ expansion.\footnote{~Recall from the discussion at the end of section~\ref{ssec:flavors} that the  massless point is well-defined at leading order in the large-$N$ limit, even though the flavor-singlet mass~\eqref{eq:flavsingm} is additively renormalized when~$k \neq 0$.} We found~$N_f + 1$ superselection sectors labeled by~$p \in \{0,1, \ldots, N_f\}$. In each sector, the~$U(N_f)$ flavor symmetry is spontaneously broken to~$U(p) \times U(N_f -p)$, as in~\eqref{eq:flavsymbreak}. The corresponding low-energy theory therefore necessarily contains massless NG bosons described by a non-linear sigma model, whose target space is the Grassmannian manifold~$\text{Gr}(p, N_f) = {U(N_f) \over U(p) \times U(N_f -p)}$ in~\eqref{eq:grassdef}. In this subsection we will describe the low-energy effective theories in the various superselection sectors in more detail. Most importantly, we will argue that effective theory not only contains massless NG bosons, but also a non-trivial Chern-Simons TQFT. 

Let us first make some comments about the effective action for the NG bosons (here we follow~\cite{Komargodski:2017keh,Freed:2017rlk}; see~\cite{DHoker:1995cvy} for general aspects of nonlinear sigma model effective actions in three dimensions):
\begin{itemize}
\item The K\"ahler form~$\omega$ of each~$\text{Gr}(p, N_f)$ target space (pulled back to spacetime) gives rise to an unbroken, topological~$U(1)$ flavor current~$\star\omega$.  There are skyrmion particles that wrap the non-trivial 2-cycle in~$\text{Gr}(p, N_f)$ (recall that~$H^2(\text{Gr}(p, N_f), \Z) = \Z$) and are therefore charged under the~$U(1)$ flavor symmetry corresponding to the current~$\star\omega$.
\item As is familiar from four dimensions~\cite{Witten:1983tw,Witten:1983tx}, the skyrmion particles described above can be identified with the baryons of~QCD$_3$, and the topological~$U(1)$ flavor symmetry can be identified with the~$U(1)_B$ baryon number (see also~\cite{Ferretti:1992fd,Ferretti:1992fga,Hong:2010sb}). This requires the inclusion of a suitable Wess-Zumino term, with quantized coefficient~$N$ (see~\cite{Komargodski:2017keh,Freed:2017rlk} for a detailed recent discussion). Among other things, this term ensures that the baryons obey the correct statistics, i.e.~they are fermions when~$N$ is odd and bosons when~$N$ is even. The Wess-Zumino term also breaks several accidental global symmetries of the low-energy NG effective theory that are not symmetries of QCD$_3$. 
\item The pion decay constant~$f^2_\pi$, which determines the overall size of the Grassmannian target manifold, scales like~$f_\pi^2  \sim N \Lambda$ in the large-$N$ limit. The~$N$-scaling follows from the fact that~$f_\pi^2$ can be extracted from two-point functions of the spontaneously broken flavor currents, which are bilinear in the quarks. Together with the Wess-Zumino term, which is also proportional to~$N$, the entire low-energy Lagrangian scales like~$N$ in the large-$N$ limit. As in four dimensions, this renders the NG effective theory semi-classical in the large-$N$ limit. 
\end{itemize}

We will now argue that the NG effective actions discussed above must generically be supplemented by a non-trivial, decoupled Chern-Simons TQFT. To this end, we revisit the mass deformations analyzed in section~\eqref{ssec:massdeform} above. If we choose the mass eigenvalues~$m_i$ as in~\eqref{eq:pposms}, so that~$p$ of them are positive, while the other~$N_f - p$ are negative, the degeneracy between the different superselection sectors is broken, and the sector labeled by~$p$ is singled out as the true vacuum sector. The mass term induces a potential on the Grassmannian NG target space~$\text{Gr}(p, N_f)$, which leads to a unique, gapped vacuum. This analysis is valid as long as~$|m_i | \ll \Lambda$. If we increase the masses beyond this regime, we may in principle encounter one or several phase transition before we reach the semi-classical asymptotic regime where~$|m_i| \gg \Lambda$. Once we reach this regime, we can reliably integrate out the massive fermions. The fact that~$p$ of them have positive masses, while the other~$N_f -p$ have negative masses, implies that the large-mass asymptotic phase is described by an~$SU(N)$ Chern-Simons theory with a shifted level,
\begin{equation}\label{eq:levelshift}
k \quad \longrightarrow \quad k + p - {N_f \over 2}~.
\end{equation}

\begin{figure}[!th]
\centerline{\includegraphics[scale=0.48]{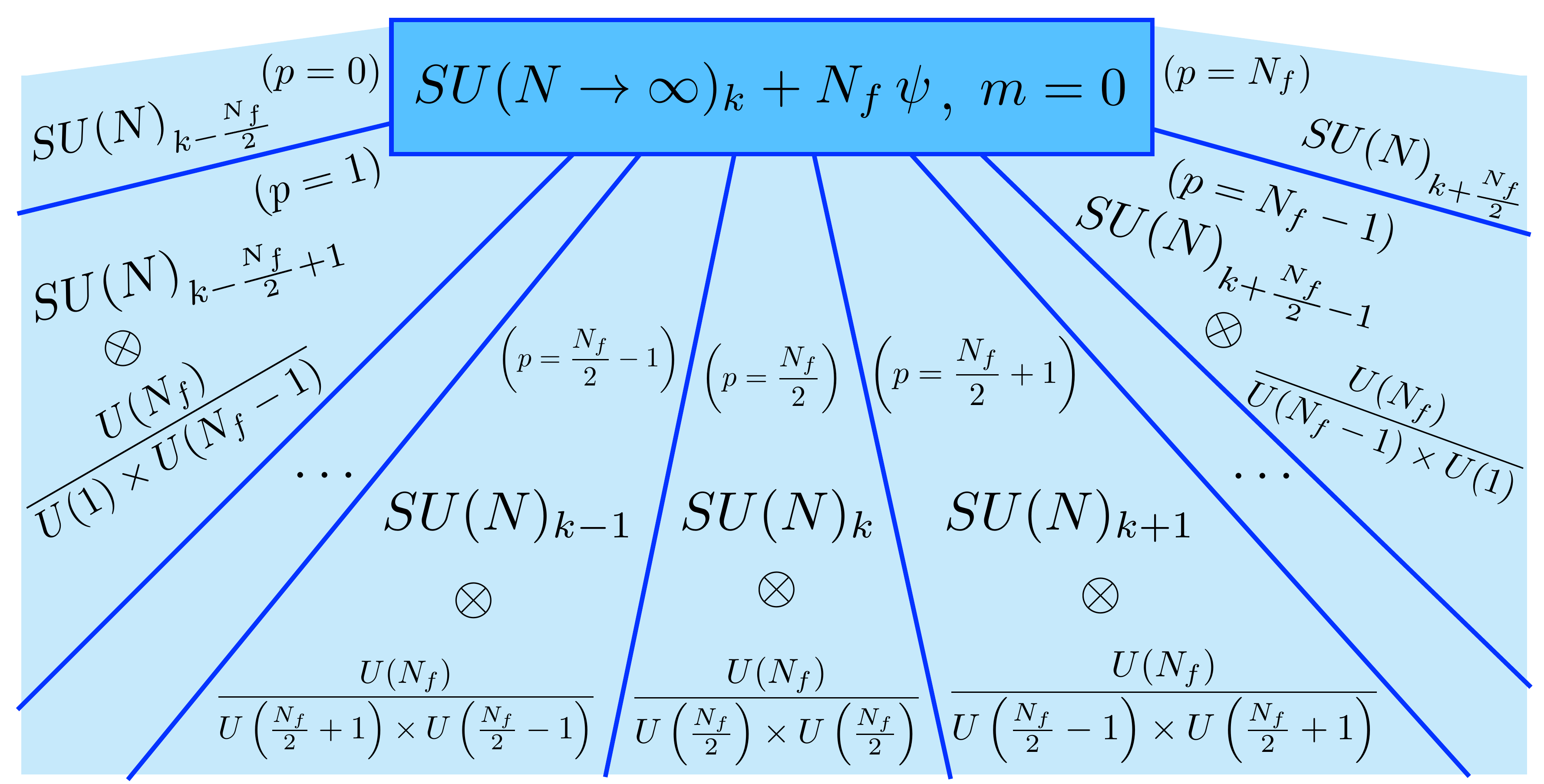}}
\caption{Degenerate superselection sectors of massless QCD$_3$ in the~$N \rightarrow \infty$ limit. For each sector we indicate the integer~$p$ appearing in~\eqref{eq:largenefts}, the low-energy Chern-Simons TQFT, and the target space for the NG bosons (i.e.~the flavor-symmetry breaking pattern). }
\label{fig:phasesstrict}
\end{figure}

In the absence of any evidence to the contrary, we make the minimal assumption that the phases at~$|m_i| \ll \Lambda$ and~$|m_i| \gg \Lambda$ are smoothly connected, i.e.~we assume that there are no phase transitions as we dial the~$m_i$ from small to large values, as long as no eigenvalue passes through zero. It then follows that the NG sigma model in each superselection sector, with target space~$\text{Gr}(p, N_f)$, is accompanied by a Chern-Simons TQFT with gauge group~$SU(N)$ and a shifted level~\eqref{eq:levelshift}. We will therefore schematically indicate the low-energy degrees of freedom in each superselection sector as follows,
\begin{equation}\label{eq:largenefts}
\text{Gr} \left(p, N_f\right) \otimes SU(N)_{k + p -{N_f \over 2}}~, \qquad  p \in \{0, 1, \ldots, N_f\}~.
\end{equation}
The different superselection sectors of massless large-$N$ QCD$_3$, and their respective low-energy degrees of freedom, are summarized in figure~\ref{fig:phasesstrict}. 

\newpage

As in section~\ref{ssec:ymcs}, the presence of the Chern-Simons TQFTs in~\eqref{eq:largenefts} has important consequences, despite being formally subleading in the large-$N$ expansion. Rather than a fully confined theory with color-singlet NG bosons and baryons, we find deconfined topological gauge fields at long distances. The Wilson lines of the TQFT in the deep IR, which descend from the Wilson lines of the QCD$_3$ theory in the UV, must end on massive degrees of freedom that carry color charge. Intuitively, these are the quarks of QCD$_3$, but as we will see in section~\ref{sec:duality}, in the large-$N$ limit it is more appropriate to think of them as dual quarks. (The fact that the IR Wilson lines should end on massive charges also follows from the fact that the microscopic QCD$_3$ theory does not have a 1-form global symmetry.) Thanks to the Chern-Simons term, these deconfined quarks  are anyons. Goldstone's theorem implies that the NG particles are necessarily exactly stable, color-neutral, spinless bosons. Hence they do not directly couple to the Chern-Simons gauge fields. By contrast, baryons can fractionalize and decay into anyonic quarks. This decay process is not captured by the low-energy effective theory described above, in which the NG sigma model and the Chern-Simons gauge fields are completely decoupled. The decay process is possible because the microscopic~$U(1)_B$ baryon charge~$B$ receives contributions from both the NG Lagrangian (via the skyrmion current~$\omega$) and from the low-energy TQFT. Since the TQFT contribution is more naturally written using level-rank dual variables, we will revisit this point in section~\ref{sec:duality}, where we discuss duality.

\section{Beyond Leading Order in the Large-$N$ Expansion}\label{sec:oneovern}

In section~\ref{sec:largenqcd3} we analyzed the vacuum structure of massless QCD$_3$ at~$\CO(N)$ in the large-$N$ limit. We found~$N_f+1$ exactly degenerate superselection sectors with different symmetry-breaking patterns, populated with NG bosons and Chern-Simons TQFTs. Turning on generic mass eigenvalues~$m_i$ singles out one of these superselection sectors, and within each sector there is a single true vacuum, which may harbor a non-trivial TQFT. Dialing any~$m_i$ through zero leads to a first-order phase transition. By contrast, dialing the masses to infinity smoothly connects with the asymptotic semi-classical regime, which can be reliably analyzed by integrating out the heavy fermions. In particular, we do not encounter any phase transitions at non-zero values of the~$m_i$.  In this section we will explore how this picture is modified at~$\CO(1)$, i.e.~at the first subleading order, in the large-$N$ expansion. As we will see, the exact degeneracy between the different superselection sectors is lifted at this order. Generically, only one superselection sector will remain, while the other~$N_f$ sectors become metastable.

\subsection{The Annulus Diagram Without Chern-Simons Terms} \label{ssec:annulus}

When we analyzed the structure of the leading~$\CO(N)$ effective potential for the meson vev~$\langle \mathbf M\rangle$ in section~\ref{ssec:disksb} using disk diagrams, there was no essential difference between theories with or without Chern-Simons terms, because their effects are subleading in the large-$N$ limit.  As explained below~\eqref{eq:rescymcs}, this is because each insertion proportional to~$k$ into a given diagram is suppressed by a factor of~$1 \over N$. Similarly, it was explained at the end of section~\ref{ssec:flavors} that the origin~$m = 0$ of the flavor-singlet mass~$\mathbf m = m \1$ in~\eqref{eq:flavsingm}  is well-defined in the large-$N$ limit, but becomes ambiguous at the next-to-leading order. For these reasons we will first analyze the~$\CO(1)$ corrections to the effective potential~$V\left(\langle \mathbf M\rangle \right)$ in QCD$_3$ with vanishing Chern-Simons level, $k = 0$, and an even number~$N_f$ of flavors (in accord with~\eqref{eq:csquant}). In this theory, the massless point~$\mathbf m = 0$ is the well-defined point in parameter space where time-reversal symmetry~$\mathsf T$ is restored. We begin by analyzing~$V\left(\langle \mathbf M\rangle\right)$ in this massless, $\mathsf T$-invariant theory. The effects of allowing for a non-zero Chern-Simons level~$k$ will be discussed in section~\ref{ssec:includecs} below. 

The~$\CO(N)$ effective potential in section~\ref{ssec:disksb} was constructed by summing disk diagrams, with a single fermion loop on the edge of the disk. This gave rise to the single-trace structure in~\eqref{eq:largenveff}, and (after following~\eqref{eq:mesmatdiag} and diagonalizing the meson vev, $\langle \mathbf M \rangle = \Lambda^2 \text{diag}\left(x_1, \ldots, x_{N_f}\right)$ with~$x_i \in \R$) the effective potential in~\eqref{eq:veffviafx}, with no interactions between the different meson eigenvalues~$x_i$,
\begin{equation}\label{eq:onveffxibis}
V(x_i)\big|_{\CO(N)} = N \Lambda^3 \sum_{i = 1}^{N_f} F(x_i)~, \qquad F(-x) = F(x)~.
\end{equation}
The fact that~$F(x)$ is an even function follows from the time-reversal symmetry~$\mathsf T$ of the~$k =0$ theory, which acts via~$\mathsf T\left(\mathbf M\right) = - \mathbf M$, and hence~$\mathsf{T}(x_i) =-x_i$ for all eigenvalues~$x_i$.

\begin{figure}[!th]
\centerline{\includegraphics[scale=0.3]{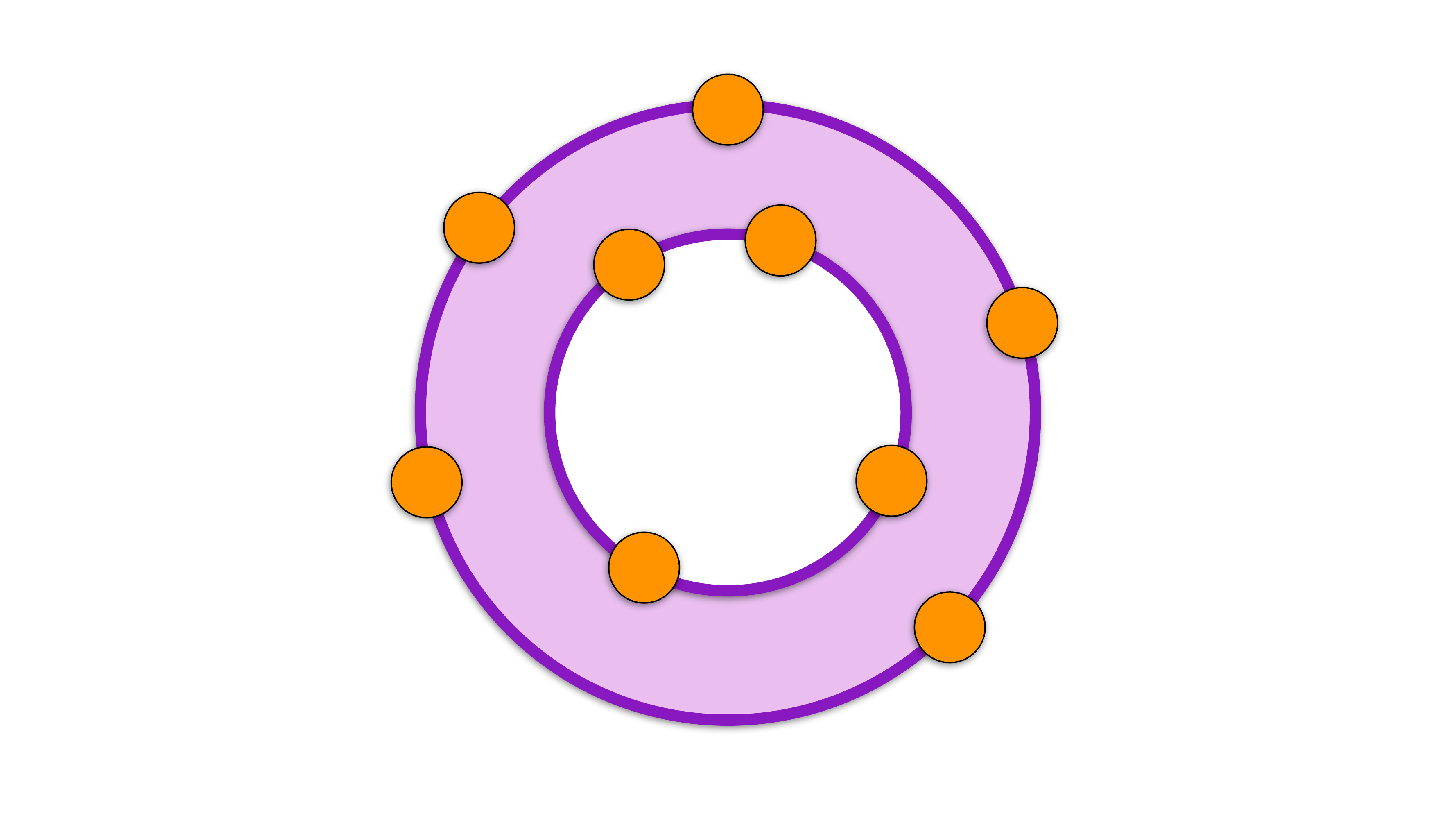}}
\caption{Annulus diagram. The orange circles on the boundary represent the insertion of meson operators.}
\label{annulus}
\end{figure}

If we wish to include next-to-leading order corrections, which scale like~$\CO(1)$ in the large-$N$ limit, we must sum over annulus diagrams, with a pair of fermion loops at the boundaries of the annulus, and insertions of the meson operator~$\mathbf M$ along these boundaries, see figure \ref{annulus}. The interior of the annulus is filled in with arbitrary planar gluon diagrams. Each such diagram leads to a double-trace contribution of the form~$ \tr\big(\langle \mathbf M\rangle^n\big) \tr\big(\langle \mathbf M\rangle^{n'}\big)$ to the effective potential, and time reversal symmetry~$\mathsf T$ requires that~$n + n' \in 2 \Z$. Such double-trace terms lead to pairwise interactions of the meson eigenvalues~$x_i$,
\begin{equation}\label{eq:gxixj}
V(x_i)\big|_{\CO(1)} = \Lambda^3 \sum_{i, j = 1}^{N_f} G\left(x_i, x_j\right)~, \qquad G\left(-x, -y\right) = G\left(x, y\right)~,
\end{equation}
where the condition $G\left(-x, -y\right) = G\left(x, y\right)$ follows from~$\mathsf T$-invariance.

Since we will not explore higher orders in the large-$N$ expansion, it is sufficient to restrict the~$\CO(1)$ annulus potential in~\eqref{eq:gxixj} to the minima of the~$\CO(N)$ disk potential in~\eqref{eq:onveffxibis}, which lie at the points~$x_i = \pm 1$. Given this restriction, we can rewrite~\eqref{eq:gxixj} in terms of the quantity~$\Delta = \half \left(G(1,1) - G(1,-1)\right)$ as follows,
\begin{equation}\label{eq:antiferro}
V\left(x_i\right)\big|_{\CO(1)} = \Lambda^3 \Delta \sum_{i, j = 1}^{N_f} x_i x_j~.
\end{equation}
Here we have dropped an immaterial constant, which does not depend on the~$x_i$. We will now argue that~$\Delta > 0$, so that the~$\CO(1)$ effective potential leads to antiferromagnetic (i.e.~repulsive) interactions between eigenvalue pairs. 

To see this, we analyze the effect of the~$\CO(1)$ potential~\eqref{eq:antiferro} on the~$N_f+1$ degenerate superselection sectors of vacua that are present at~$\CO(N)$ for either sign of~$\Delta$:
\begin{itemize}
\item[a)] If~$\Delta < 0$, the superselection sectors of lowest energy maximize the number of aligned eigenvalue pairs. This happens when~$p = 0$, so that all~$x_i = 1$, or when~$p = N_f$, so that all~$x_i = -1$. In either case the~$U(N_f)$ flavor symmetry is unbroken, but time-reversal symmetry~$\mathsf T$ is spontaneously broken. Therefore the superselection sectors with~$p = 0$ and~$p = N_f$, which are exchanged by~$\mathsf T$, remain exactly degenerate. 
\item[b)] If~$\Delta > 0$, the superselection sector of lowest energy maximizes the number of anti-aligned eigenvalue pairs. This happens when~$p = {N_f \over 2}$, so that~${N_f \over 2}$ eigenvalues are at~$x_i = 1$, while the other~${N_f \over 2}$ eigenvalues are at~$x_i = -1$, leading to the symmetry-breaking pattern~$U(N_f) \; \rightarrow \; U\left({N_f \over 2}\right) \times U\left({N_f \over 2}\right)$. Time-reversal symmetry~$\mathsf T$ maps this superselection sector to itself. 
\end{itemize} 
As we reviewed around~\eqref{eq:vwmasses}, it was shown in~\cite{Vafa:1983tf,Vafa:1984xh} that massless~QCD$_3$ with~$k = 0$ and even~$N_f$ necessarily contains a~$U\left({N_f \over 2}\right) \times U\left({N_f \over 2}\right)$-preserving superselection sector that maps to itself under~$\mathsf T$. (This superselection sector can be identified by turning on suitably symmetric mass terms and taking the zero-mass limit.) This is only consistent with the preceding discussion if option~b) above is realized, i.e.~if~$\Delta > 0$. 

A different argument that~$\Delta > 0$, that we now briefly sketch, is based on the observation that the splitting~$\Delta \sim G(1,1)-G(1,-1)$ only receives contributions from effective-potential terms of the form~$ \tr\big(\langle \mathbf M\rangle^n\big) \tr\big(\langle \mathbf M\rangle^{n'}\big)$ with~$n, n'$ both odd. (Terms with~$n, n'$ both even contribute equally to~$G(1, 1)$ and~$G(1,-1)$.) Such terms can be thought of as arising from a closed-string process. More precisely, they arise from~$\mathsf T$-odd scalar glueballs propagating from one boundary of the annulus diagram to the other.\footnote{~In~$2+1$ dimensions, only massive scalar particles can be further classified as~$\mathsf T$-even or~$\mathsf T$-odd. We are grateful to Simon Caron-Huot for a related discussion.} (By contrast, terms with even~$n, n'$ only receive contributions from intermediate~$\mathsf T$-even scalar glueballs.) Using the representation of this process as a sum over intermediate glueball states, it can be argued that its coefficient, and hence~$\Delta$, should be positive.

The upshot of the preceding discussion is that the coefficient~$\Delta$ in~\eqref{eq:antiferro} is positive, $\Delta > 0$, and this in turn singles out the~$p ={N_f \over 2}$ superselection sector as the true vacuum up to and including~$\CO(1)$ in the large-$N$ expansion. Since this vacuum is degenerate with the other~$N_f$ superselection sectors (where~$p \neq {N_f \over 2}$) at~$\CO(N)$, there are~$N_f$ metastable vacuum sectors, which are split from the true vacuum by an~$\CO(1)$ energy difference. Since no degeneracy remains once we include the~$\CO(1)$ correction from the annulus, and since the higher order contributions in the large $N$ expansion decay as $N\to\infty$,  it is plausible that our qualitative conclusions remain correct at large but finite~$N$. In particular, we expect the theory at~$N \gg 1$ to have metastable vacua and to display associated first-order phase transitions. Below we will explicitly exhibit these transitions. 

\subsection{A Sequence of First-Order Phase Transitions}\label{ssec:firstorder}

We will now study the phase diagram of the~$k = 0$ theory as a function of the flavor-singlet mass deformation~$\mathbf m = m \1$ in~\eqref{eq:flavsingm}. (As in section~\ref{ssec:massdeform} it is straightforward to consider more general mass deformations.) The~$N_f$ metastable vacua discussed at the end of the previous subsection will reappear in this context. 

For small~$m$ (i.e.~$m \ll \Lambda$) we proceed as in section~\ref{ssec:massdeform} and add the first-order mass deformation~\eqref{eq:delveigenv} to the effective potential of the massless theory. As we will self-consistently confirm below, this approximation is sufficient to reliably capture the effects we describe here. We must therefore minimize the sum of the~$\CO(1)$ effective potential~\eqref{eq:antiferro} and the mass deformation~\eqref{eq:delveigenv},
\begin{equation}\label{eq:totveff}
V_\text{tot.}(x_i) = \Lambda^3 \Delta \sum_{i, j=1}^{N_f} x_i x_j + N m \Lambda^2 \sum_{i = 1}^{N_f}  x_i~, \qquad \Delta > 0~, \qquad x_i = \pm 1~,
\end{equation}
while the~$\CO(N)$ effective potential provides the constraint~$x_i = \pm 1$. Note that the potential in~\eqref{eq:totveff} effectively describes a classical antiferromagnet in an external magnetic field~$\sim N m$. 

In the superselection sector labeled by $p \in \left\{0, 1, \ldots, N_f\right\}$, the first~$p$ eigenvalues reside at~$x_i = -1$, while the remaining~$N_f-p$ eigenvalues reside at~$x_i =1$ (see~\eqref{eq:pvacdef}). It is straightforward to evaluate the potential~\eqref{eq:totveff} in every such superselection sector, 
\begin{equation}\label{eq:vofp}
V_\text{tot.}(p) = \Lambda^3 \Delta \left(N_f -2p\right)  \left(N_f - 2 p  + {Nm \over \Lambda \Delta}\right)~.
\end{equation}
The integer~$p = p_*(m) \in \left\{0, 1, \ldots, N_f\right\}$ that labels the true vacuum sector is obtained by minimizing~\eqref{eq:vofp}. This would straightforward if~$p$ were a continuous variable. Instead, we use~$[\![ x ]\!]$ to denote the integer nearest to~$x$. We will only use this notation if~$x$ is not a half-integer (see below). Then the vacuum sector is labeled by
\begin{equation}\label{eq:vacpstar}
p_*(m) = {N_f \over 2} + \left[\!\!\left[ {N m \over 4 \Lambda \Delta} \right]\!\!\right]~.
\end{equation}
Setting~$m = 0$, we recover the result~$p_*(0) = {N_f \over 2}$ of section~\ref{ssec:annulus} above. 

\begin{figure}[!th]
\centerline{\includegraphics[scale=0.8]{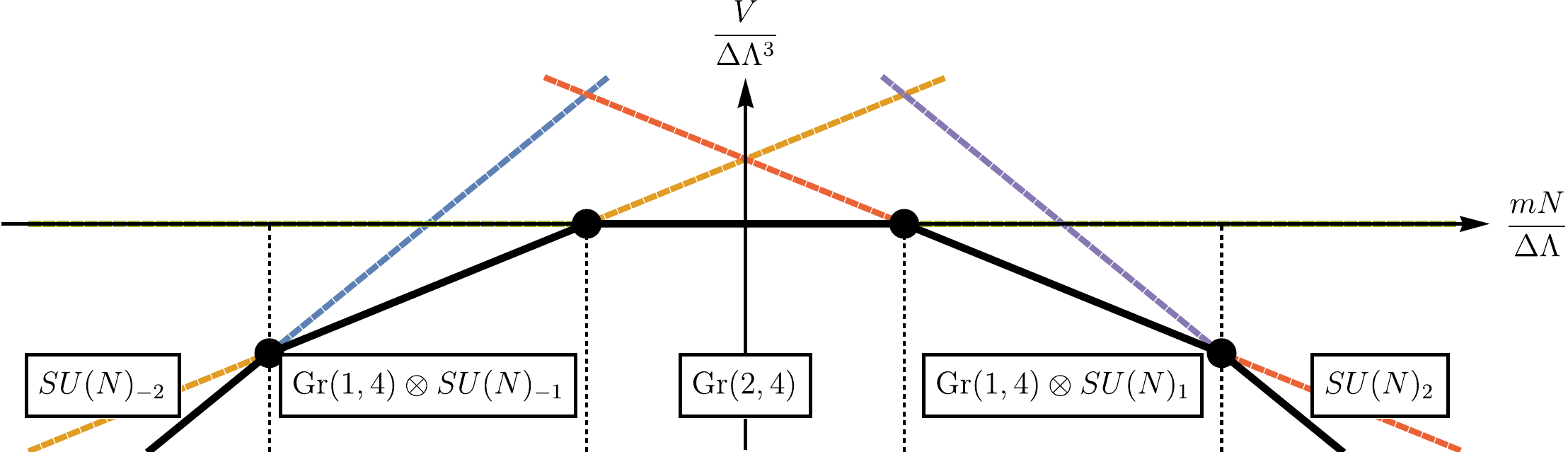}}
\bigskip
\bigskip
\caption{The potential \eqref{eq:vofp} as a function of $m$ in the different superselection sectors for $N_f=4$. First-order phase transitions are indicated by black dots. The black line represents the true vacuum.}
\label{PhasesDiagram}
\end{figure}

If we dial~$m$ away from~$0$, the function~$p_*(m)$ in~\eqref{eq:vacpstar} will jump by~$\pm 1$ whenever ${N m \over 4 \Lambda \Delta}\in \Z + \half$ lies exactly between two integers. This happens when~${N m \over \Lambda \Delta} \equiv 2\ {\rm mod}\ 4$. At these points two adjacent superselection sectors (with consecutive values of~$p$) are exactly degenerate, and there is a first-order phase transition from one sector to the other. For instance, the first such transition point for positive~$m$ occurs when~${N m \over \Lambda \Delta} = 2$, where the superselection sectors with~$p = {N_f \over 2}$ and~$p = {N_f \over 2} + 1$ are exactly degenerate. If we dial~${N m \over \Lambda \Delta}$ through this point, we jump from the former to the latter sector. As we dial~$m$ from large negative to large positive values, we therefore traverse all superselection sectors we found in section~\ref{sec:largenqcd3}, labeled by~$p \in \left\{0, 1, \ldots, N_f\right\}$, in consecutive order. Rather than being exactly degenerate, as was the case at leading order in the large-$N$ expansion, the different sectors are now separated by a sequence of first-order phase transitions in the mass parameter~$m$. An example of the phase diagram (for the case~$N_f = 4$) as a function of~$m$, as well as the low-energy degrees of freedom in each phase, appears in figure~\ref{PhasesDiagram}.

Once we reach the asymptotic pure Chern-Simons phases without NG bosons, which correspond to~$p = 0$ and~$p = N_f$, there are no additional transitions if we further increase~$|m|$. The transition into the asymptotic phases occurs when~${N m \over \Lambda \Delta} = \pm \left(2 N_f - 2\right)$, which is congruent to~$2~(\text{mod}~4)$ since~$N_f$ is even. The fact that all first-order transitions discussed above, including the ones to the asymptotic phases, occur when the mass parameter~$m$ is of order~$|m| \sim {\Lambda \over N}$ retroactively justifies retaining only the linear~$\CO(m)$ term in the effective potential~\eqref{eq:totveff}, as well as using the~$O(N)$ effective potential to constrain the eigenvalues to $\pm1$.

\subsection{Incorporating the Chern-Simons Term}\label{ssec:includecs}

So far we have only discussed~QCD$_3$ with~$k = 0$ at~$\CO(1)$ in the large-$N$ expansion. Here we summarize the necessary modifications for non-vanishing Chern-Simons level~$k$. Now~$k$ makes a direct appearance in the effective potential~$V\left(\langle \mathbf M \rangle\right)$ for the meson vev~$\langle \mathbf M\rangle$, via disk diagrams diagrams involving a single Chern-Simons propagator or vertex. As explained below~\eqref{eq:rescymcs}, such diagrams are suppressed by a factor~$\sim {k \over N}$ relative to the leading~$\CO(N)$ disk diagrams, and hence such a disk diagram would give a contribution proportional to~$k$. Given that the~$\CO(N)$ effective potential restricts the meson eigenvalues to~$x_i = \pm 1$, the only possible effect of the Chern-Simons level is therefore to contribute a term~$\sim k \Lambda^3 \sum_i x_i$ to~$V\left(\langle \mathbf M \rangle\right)$. 

Such a term takes exactly the same form as the contribution~$N m \Lambda^2 \sum_i x_i$ of the flavor-singlet mass~$\mathbf m = m \1$ in~\eqref{eq:totveff}. This is consistent with the discussion at the end of section~\ref{ssec:flavors}, according to which the presence of a non-zero~$k$ additively renormalizes the flavor-singlet mass~$m$ by an amount~$\Delta m \sim {k \Lambda \over N}$. We can therefore simply absorb the contribution of~$k$ to the~$\CO(1)$ effective potential into a redefinition of the flavor-singlet mass~$m$ and study the phase diagram as a function of that mass. Note that the origin~$m = 0$ is no longer a distinguished point in parameter space, because the presence of~$k$ explicitly breaks time-reversal symmetry~$\mathsf T$. 

With these caveats, the phase diagram of the~$k \neq 0$ theory as a function of~$m$ is nearly identical to that of the~$k = 0$ theory described in section~\ref{ssec:firstorder}. As~$m$ is varied from~$-\infty$ to~$\infty$, there is a sequence of first-order phase transition traversing all superselection sectors found in section~\ref{sec:largenqcd3}, which are labeled by~$p \in \left\{0, 1, \ldots, N_f\right\}$, in consecutive order. The only difference is that the levels of the Chern-Simons TQFTs in the different superselection sectors are shifted by the bare level~$k$. As we emphasized in the introduction, this picture does not depend on whether~$k$ is larger or smaller than~$N_f \over 2$. However, recall that when $ 0 \leq k < {N_f \over 2}$, the superselection sector labeled by~$p = {N_f \over 2} - k$ does not contain any low-energy Chern-Simons TQFT.  This sector describes the quantum phase studied in~\cite{Komargodski:2017keh}. By contrast, for $k\geq {N_f \over 2}$ all superselection sectors contain a nontrivial TQFT at low energies. We will see below that this fact is rather important for duality.

\section{Dualities for~QCD$_3$ in the Large-$N$ Limit}\label{sec:duality}

So far we have analyzed the large-$N$ dynamics of QCD$_3$, i.e.~$SU(N)_k$ Yang-Mills-Chern-Simons theory with~$N_f$ fundamental quarks. In this section we will examine the results of the preceding sections through the lens of duality. (For an overview of duality in the context of QCD$_3$, with references, see section~\ref{sec:intro}). Before we proceed, an important conceptual comment is in order. Three-dimensional boson-fermion dualities are believed to be infrared dualities, and hence they are cleanest, and most powerful, near a second-order phase transition described by a CFT. In our analysis of large-$N$ QCD$_3$ we have found a sequence of first-order transitions between phases containing NG bosons and Chern-Simons TQFTs. Therefore, the boson-fermion dualities are not as powerful as they would have been if the transitions were second order. Nevertheless, the dualities can correctly capture the sequence of first-order phase transitions, the nonlinear sigma models that describe the gapless NG bosons, and the low-energy Chern-Simons theories, as well as their couplings to background fields. (It would be nice to check if the theories on the domain walls at the first-order transition points also agree.) Thus the boson-fermion dualities are informative, even though the transitions are first order.

The dual theories that arise for QCD$_3$ are~$U(\t N)_{\t k}$ Chern-Simons theories with~$N_f$ scalars in the fundamental representation of~$U(\t N)$. The rank~$\t N$ of the dual gauge group depends on~$N_f$ and~$k$, but not on~$N$. It therefore remains fixed in the large-$N$ limit. By contrast, the dual Chern-Simons level~$\t k \sim N$ becomes large, and hence the bosonic dual theories are weakly coupled at large~$N$. (Recall the discussion about weakly coupled limits of three-dimensional gauge theories in section~\ref{sec:intro}.) In this way the duality beautifully and explicitly realizes the idea that the large-$N$ limit of QCD should be a weak-coupling limit. Such gauge theories at large Chern-Simons level often flow to weakly coupled conformal fixed points. To reproduce the behavior of the fermionic theory we will therefore have to carefully adjust the scalar potential.

Indeed, the bosonic dual theories are able to succinctly describe the phases and first-order phase transitions found in sections~\ref{sec:largenqcd3} and~\ref{sec:oneovern} above, as long as the scalar potentials of these bosonic theories take a particular form that we describe in detail. This form is not compatible with the quartic scalar potentials assumed in many previous studies. 

\subsection{The Bosonic Dual Theories}\label{ssec:bosduals}

The bosonic dual theories that arise for QCD$_3$ are~$U(\t N)_{\t k}$ Chern-Simons gauge theories. The rank~$\t N = \half N_f \pm k$ of the dual gauge group and the dual Chern-Simons level~$\t k = \pm N$ are determined by the parameters~$N, k, N_f$ of QCD$_3$. Note that~$\t N$ does not depend on the number of colors~$N$, and hence remains fixed in the large-$N$ limit, while~$\t k = \pm N$ becomes large. As will be reviewed in section~\ref{ssec:oldpot} below, there are sometimes two bosonic duals with different values of~$\t N$ and~$\t k$ (corresponding to the~$\pm$ signs in the preceding formulas) for a given QCD$_3$ theory with fixed~$N, k, N_f$. The~$U(\t N)_{\t k}$ Chern-Simons theories are coupled to~$N_f$ scalars~$\phi^{a i}$ in the fundamental representation of the~$U(\t N)$ gauge group. Here~$a = 1, \ldots, \t N$ is a fundamental gauge index, while~$i = 1, \ldots, N_f$ is a fundamental index for the~$SU(N_f)$ flavor symmetry that acts on the~$\phi^{ai}$. There is also a topological~$U(1)$ flavor symmetry associated with the center of the~$U(\t N)$ gauge group, which acts on monopole operators and dynamical vortices. It is identified with the~$U(1)_B$ baryon number symmetry of QCD$_3$. 
The bosonic dual theory is described by the following Lagrangian,
\begin{equation}\label{eq:duallag}
\SL = N \left(D^\mu \phi\right)^\dagger_{a i} D_\mu \phi^{ai} \pm {N \over 4 \pi} \Tr \left( C \wedge d C + {2 \over 3} C \wedge C \wedge C \right)+ V\left(\phi\right)~.
\end{equation}
Here~$C$ is the dual~$U(\t N)$ gauge field and~$\Tr$ denotes a trace over gauge indices. (The conventions are as in section~\ref{ssec:ymcs}.)  We would like to make several comments about this Lagrangian:
\begin{itemize}
\item The~$U(1)_B$ baryon charge~$B$ is given by the following topological charge,
\begin{equation}\label{eq:dualmicroB}
B={1 \over 2 \pi} \int_{{\cal M}_2}\Tr dC~,
\end{equation}
i.e. by the integral of the first Chern class of the~$U(\t N)$ gauge field over a 2-cycle.
\item If we include a Yang-Mills kinetic term~$ {1 \over 4 g^2_C} \Tr \left(F_C \wedge \star F_C\right)$ for the dual gauge field~$C$, the Chern-Simons term (with level~$\t k = \pm N$) will induce a gluon mass~$m_\text{gluon} \sim N g_C^2$. The dual~$U(\t N)$ gauge group, and hence its gauge coupling~$g_C$, are held fixed in the large-$N$ limit, so that the gluon mass is~$\CO(N)$ at large~$N$. Since we do not in general expect the dual theory to correctly capture such massive states,  we can integrate out the massive gluons and omit the Yang-Mills term for~$C$. Baryons are visible on both sides of the duality, despite the fact that their masses are also~$\CO(N)$ in the large-$N$ limit~\cite{Witten:1979kh}. In the bosonic duals they are described by magnetic vortices of the dual~$U(\t N)$ gauge theory.

\item For future convenience, we have normalized the scalar fields~$\phi^{a i}$ so that their kinetic terms contains an explicit factor of~$N$. This will help us clarify the weak coupling nature of the large-$N$ limit. 

\item In many previous studies, the scalar potential~$V(\phi)$ of the bosonic dual theory was taken to be a quartic polynomial in~$\phi$. More recently the authors of~\cite{Aharony:2018pjn} considered bosonic duals with sextic potentials. As we will discuss in section~\ref{ssec:dualsp} below, the dual scalar potential must have certain properties in order to capture the phases of large-$N$ QCD$_3$. These properties are not compatible with a quartic potential, but they can be achieved with a sextic potential (or more general higher-order terms, see below). This is not entirely surprising: when the theory flows to a nontrivial fixed point describing a second-order phase transition, the sextic term may be an irrelevant operator. If this is the case, it can be safely omitted from the discussion. By contrast, if the phase transitions are first order, not all field vevs need to become suitably small as we approach the transition, so that higher-order polynomial corrections to the potential are generally important. Such corrections can be viewed as dangerously irrelevant.
\end{itemize}

The complex scalars~$\phi^{ai}$ can be organized into a~$\t N \times N_f$ matrix, with~$\t N$ rows indexed by a gauge index~$a$ and~$N_f$ columns indexed by a flavor index~$i$. We can define gauge-invariant dual meson operators~${{\CM}_i}^j$,\footnote{~Recall the definition of the fermion meson operator~${M_i}^j$ from~\eqref{eq:mesondef}, ${M_i}^j = {1 \over N} \b \psi_i \psi^j$.}
\begin{equation}\label{eq:bosmesdef}
 {\CM_i}^j = \phi^\dagger_{a i} \phi^{a j}~.
\end{equation}
Note that~${\CM_i}^j$ is a positive-definite hermitian~$N_f \times N_f$ matrix, which transforms in the adjoint representation of the~$SU(N_f)$ flavor symmetry. Using~$U(\t N)$ gauge and~$SU(N_f)$ flavor transformations, the matrix~$\phi^{ai}$ can be put into diagonal form. For this purpose we distinguish the cases~$\t N \geq N_f$ and~$\t N < N_f$:\begin{itemize}
\item[1.)] If~$\t N \geq N_f$, then~$\phi^{ai}$ can be expressed in terms of~$N_f$ non-negative singular values~$y_i$ (with~$i = 1, \cdots, N_f$), which we are free to arrange in descending order,
\begin{equation}\label{eq:phibign}
\phi^{ai} = \begin{pmatrix}
y_1 & 0 &  \cdots & 0 \\
0 & y_2 & \ddots & \vdots  \\
\vdots & \ddots & \ddots & 0 \\
0 & \cdots & 0 &  y_{N_f} \\
0 & 0 & \cdots & 0\\
\vdots & \vdots & \ddots & \vdots \\ 
0 & 0 & \cdots & 0 
\end{pmatrix} = \begin{pmatrix} \text{diag}\left(y_1, \ldots, y_{N_f}\right) \\ \mathds{O}_{\left(\t N - N_f\right) \times N_f} \end{pmatrix}~, \qquad y_1 \geq y_2 \geq \cdots \geq y_{N_f} \geq 0~.
\end{equation}
Here~$\mathds{O}_{m \times n}$ denotes an~$m$ by~$n$ matrix whose entries all vanish. The rank of~$\phi^{ai}$ is the number of non-vanishing~$y_i$, and the largest possible rank is~$N_f$. Consequently the dual meson operator~${\CM_i}^j$ in~\eqref{eq:bosmesdef}, whose maximal rank is also~$N_f$, is given by
\begin{equation}\label{eq:meslargen}
{\CM_i}^j = \text{diag} \left(y_1^2 \, , \, \ldots \, , \, y_{N_f}^2\right)~.
\end{equation}

\item[2.)] If~$\t N < N_f$, then~$\phi^{ai}$ can be expressed in terms of~$\t N$ non-negative singular values~$y_a$ (with~$a = 1, \cdots, \t N$), which we again arrange in descending order,
\begin{equation}
\begin{split}\label{eq:phismalln}
\phi^{ai} = \begin{pmatrix}
y_1 & 0 & \cdots & \cdots & 0 &  \cdots & 0\\
0 & y_2 & 0 & \cdots & 0 & \cdots & 0 \\
\vdots & \ddots & \ddots  & \ddots & \vdots  & \ddots &  \vdots \\
0 & \cdots & 0 & y_{\t N} & 0 & \cdots & 0
\end{pmatrix} =  & \begin{pmatrix} \text{diag}\left(y_1, \ldots, y_{\t N}\right) \, , \, \mathds{O}_{\t N \times \left(N_f - \t N\right)} \end{pmatrix}~,\\
& \hskip55pt y_1 \geq y_2 \geq \cdots \geq y_{\t N} \geq 0~.
\end{split}
\end{equation}
Now the maximal rank of~$\phi^{ai}$ and~${\CM_i}^j$ is~$\t N$, and the dual meson operator necessarily has at least~$N_f - \t N$ zero eigenvalues,
\begin{equation}\label{eq:messmalln}
{\CM_i}^j = \text{diag}\Big(y_1^2 \, , \, \ldots \, , \,  y_{\t N}^2 \, , \overbrace{\, 0, \, \ldots. \, , 0 \,}^{N_f -\t N} \Big)~.
\end{equation}

\end{itemize}
\noindent The parametrizations~\eqref{eq:phibign}, \eqref{eq:phismalln} for~$\phi^{ai}$ and~\eqref{eq:meslargen}, \eqref{eq:messmalln} for~${\CM_i}^j$ will facilitate our discussion of higgsing and flavor-symmetry breaking below. 

\subsection{Duality Scenarios with Quartic Scalar Potentials}\label{ssec:oldpot}

As we reviewed in section~\ref{sec:intro}, previous work on the phases of~QCD$_3$ uncovered two qualitatively different parameter regimes: $k \geq \half N_f$, and~$k  < \half N_f$. We will now review the proposed bosonic duals that cover these two regimes (see section~\ref{sec:intro} for references). We offer some incidental new observations about the structure of quartic potentials in these scenarios. Similar comments appear in~\cite{Argurio:2019tvw}.

\subsubsection{$k \geq {N_f \over 2 }$} \label{sssec:largek}

When~$k \geq {N_f \over 2 }$, there is a proposed bosonic dual of the kind reviewed in section~\ref{ssec:bosduals}, with Lagrangian~\eqref{eq:duallag}. It is a~$U(\t N)_{\t k}$ gauge theory with~$N_f$ fundamental bosonic flavors~$\phi^{ai}$. The dual gauge group has rank~$\t N = k + {N_f \over 2}$, and the dual Chern-Simons level is~$\t k = -N$. Since~$\t N \geq N_f$, we are in the regime discussed around~\eqref{eq:phibign}. In short, the bosonic dual is
\begin{equation}\label{eq:onedual}
U\left(\t N = k + {N_f \over 2 }  \geq N_f\right)_{\t k = -N}+ \phi^{ai} \qquad \big(a = 1, \ldots, \t N \;; \;  i = 1, \ldots, N_f\big)~.
\end{equation}
In previous studies, this dual was assumed (either explicitly or implicitly) to be equipped with a certain quartic scalar potential~$V_4(\phi)$. By gauge invariance, this potential~$V_4(\phi)$ is a function of the dual meson~${\CM_i}^j$ defined in~\eqref{eq:bosmesdef}. Moreover, it must respect the~$SU(N_f)$ flavor symmetry of the theory, and hence it is a sum of products of traces~$\sim \tr \left(\CM^n\right)$. (As before, we use $\tr$ to denote a trace over flavor indices.) In general, both single- and multi-trace terms can appear. The most general quartic potential~$V_4(\phi)$ takes the form
\begin{equation}\label{eq:quartpot}
V_4(\phi) = \mu^2 \tr \left(\CM\right) + \lambda_1 \tr \left(\CM^2\right) + \lambda_2 \tr\left(\CM\right)^2~.
\end{equation}
Here~$\lambda_1$ and~$\lambda_2$ respectively denote the single- and double-trace quartic couples, and the mass-squared~$\mu^2$ can be either positive or negative. 

In order to analyze the potential in~\eqref{eq:quartpot}, we use its~$U(\t N) \times SU(N_f)$ invariance to parametrize~$\phi^{ia}$ in terms of~$N_f$ singular values~$y_i$, as in~\eqref{eq:phibign}, so that~${\CM_i}^j = \text{diag}\left(y_1^2, \ldots, y_{N_f}^2\right)$,
\begin{equation}\label{eq:v4y}
V_4(\phi) = \mu^2 \sum_{i =1}^{N_f} y_i^2 + \lambda_1 \sum_{i = 1}^{N_f} y_i^4 + \lambda_2 \left(\sum_{i =1}^{N_f} y_i^2\right)^2~.
\end{equation}
By examining the behavior for large, positive~$y_i$, we can obtain various stability bounds on~$\lambda_{1,2}$. For instance, by examining the slice~$y_1 = \cdots = y_p  =  t$ and~$y_{p + 1} = \cdots = y_{N_f} = 0$, we find that~$V_4(\phi) \sim p \left(\lambda_1 + p \lambda_2\right) t^4$ for large, positive~$t$. This leads to the bound
\begin{equation}\label{eq:stabcons}
\lambda_1 + p \lambda_2 \geq 0~, \qquad p = 1, \ldots, N_f~.
\end{equation}
If~$\lambda_2 \geq 0$, then all constraints in~\eqref{eq:stabcons} follow from~$\lambda_1 + \lambda_2 \geq 0$; by contrast, if~$\lambda_2 < 0$, all other constraints follow from~$\lambda_1 + N_f \lambda_2 \geq 0$.  

Let us analyze the classical vacua, i.e.~the minima of the potential~\eqref{eq:v4y}. Its first derivatives vanish at points~$y_i$ that satisfy
\begin{equation}\label{eq:yext}
y_i = 0 \qquad \text{or} \qquad {\mu^2 \over 2} + \lambda_1 y_i^2 + \lambda_2 \sum_{j = 1}^{N_f} y_j^2 = 0~.
\end{equation}
By examining the quadratic terms in~\eqref{eq:v4y} we see that the point~$y_i = 0$ is a minimum when~$\mu^2 \geq 0$ and a maximum when~$\mu^2 < 0$. Summing over the free index~$i$ in the second equation of~\eqref{eq:yext} for all non-zero~$y_i$ (i.e.~from~$i =1$ to~$i = \text{rank}(\CM)$) leads to\footnote{~We are grateful to the authors of~\cite{Argurio:2019tvw} for a useful discussion about the case with non-maximal rank.}
\begin{equation}\label{eq:ysqrad}
\sum_{i = 1}^{N_f} y_i^2= - {\text{rank}(\CM) \mu^2 \over 2 \left(\lambda_1 + \text{rank}(\CM) \lambda_2\right)}~. 
\end{equation}
Due to the stability bounds~\eqref{eq:stabcons}, this only has solutions away from the origin when $\mu^2 < 0$. Substituting back into~\eqref{eq:yext}, we find two scenarios:
\begin{itemize}
\item If~$\mu^2 < 0$ and~$\lambda_1 \neq 0$, then for every possible value of~$\text{rank}(\CM)$, the potential has a stationary point at
\begin{equation}\label{eq:yvevgen}
y_1 = \cdots = y_{\text{rank}(\CM)} = \left(- {\mu^2 \over 2 \left(\lambda_1 + \text{rank}(\CM) \lambda_2\right)}\right)^\half~, \quad y_{\text{rank}(\CM) + 1} = \cdots = y_{N_f} = 0~.
\end{equation}
Substituting back into the potential~\eqref{eq:v4y}, we find
\begin{equation}\label{eq:v4rank}
V_4(\phi) = - {\text{rank}(\CM) \mu^4 \over 4 \left(\lambda_1 + \text{rank}(\CM) \lambda_2\right)}~, \qquad \text{rank}(\CM) = 0, 1, \ldots, N_f~.
\end{equation}
In order to determine the true minimum of~\eqref{eq:v4rank}, we must distinguish two sub-cases:
\begin{itemize}
\item[a)] If~$\lambda_1 > 0$, we can use the stability bounds~\eqref{eq:stabcons} to show that the potential is minimized when~$\CM$ has maximal rank, i.e.~when $\text{rank}(\CM) = N_f$. 
\item[b)] If~$\lambda_1 < 0$, we can use the stability bounds~\eqref{eq:stabcons} to show that the potential is minimized when~$\CM$ has minimal non-vanishing rank, i.e.~for~$\text{rank}(\CM) = 1$. 
\end{itemize}

\item If~$\mu^2 < 0$ and~$\lambda_1 = 0$, then there is a manifold of exactly degenerate classical vacua~$y_i$. It follows from~\eqref{eq:yext} that these vacua lie on an~$S^{N_f-1}$ defined by~$\sum_{i =1}^{N_f} y_i^2 = -{ \mu^2 \over 2 \lambda_2}$. In particular~$\text{rank}(\CM)$ can take any value between~$1$ and~$N_f$.
\end{itemize}

 As we already discussed in the introduction, and will review below, the existing duality proposals with quartic potentials always assume maximal color-flavor locking in the Higgs phase, which requires~$\CM$ to have maximal rank, i.e.~$\text{rank}(\CM) = N_f$. By comparing with the above analysis, we see that this can only be guaranteed if~$\lambda_1 > 0$, and we will therefore assume that this inequality holds. By contrast, $\lambda_2$ can have either sign (or vanish) as long as the stability bounds~\eqref{eq:stabcons} are satisfied. 

Note that the vacua away from the origin approach the origin as~$\mu^2 \rightarrow 0^-$. Classically, the point~$\mu^2 = 0$ therefore corresponds to a second order phase transition, where new massless particles appear. Most of the recent literature (see~\cite{Aharony:2018pjn} for an exception) assumes that only the mass parameter~$\mu^2$ should be dialed, while the quartic couplings~$\lambda_1$ and~$\lambda_2$ are assumed to be suitably generic (subject to the various positivity bounds discussed above), and hence irrelevant for the phase structure and dynamics in the deep IR. Moreover, it is usually assumed that the infrared phases are correctly captured by the classical scalar potential analyzed above, and that the phase structure is not spoiled by quantum effects. By contrast, it is well known that quantum effects can change the order of a phase transition, as in 
\cite{Coleman:1973jx,Halperin:1973jh} where quantum corrections turn a classical second-order transition into a first-order transition.

With these comments in mind, we can summarize the proposed IR behavior of~\eqref{eq:onedual}, which is claimed to be dual to the IR behavior of QCD$_3$ with~$k \geq {N_f \over 2 }$:
\begin{itemize}
\item When~$\mu^2 > 0$ we are in a phase with~$y_i = 0$, and hence~$\phi^{ai} = 0$. Therefore neither the gauge nor the flavor symmetries are spontaneously broken, and all scalars acquire a mass. Integrating them out, we are left with a~$U\left(k + {N_f \over 2 } \right)_{-N}$ Chern-Simons TQFT, which is level-rank dual to 
\begin{equation}\label{eq:irtqftone}
U\left(k + {N_f \over 2 } \right)_{-N} \qquad \longleftrightarrow \qquad SU(N)_{k + {N_f \over 2 }}~.
\end{equation}
\item When~$\mu^2 < 0$ we are in a phase where all~$y_i$ are equal and non-vanishing. Comparing with~\eqref{eq:phibign}, we see that the~$\t N \times N_f$ matrix~$\phi^{ai}$ is given by
\begin{equation}
\phi^{ai} \sim \begin{pmatrix} \1_{N_f \times N_f} \\ \mathds{O}_{(\t N - N_f) \times N_f}\end{pmatrix}~.
\end{equation}
Therefore the flavor symmetry is unbroken (upon mixing with the gauge symmetry), but the gauge group is higgsed to~$U\left(\t N - N_f\right) = U\left(k - {N_f \over 2 }\right)$. All scalars acquire a mass and, using level-rank duality, the TQFT in the deep IR is
\begin{equation}
U\left(k - {N_f \over 2}\right)_{-N}  \qquad \longleftrightarrow \qquad SU(N)_{k - {N_f \over 2}}~.
\end{equation}
\end{itemize} 
As we reviewed in section~\ref{sec:intro}, the two phases above coincide with the asymptotic phases of QCD$_3$ if we identify~$\mu^2 \sim m \Lambda$ (up to an additive constant). Here~$m$ is the flavor-singlet mass deformation of~QCD$_3$, and~$\Lambda$ is its strong-coupling scale. As we have seen in sections~\ref{sec:largenqcd3} and~\ref{sec:oneovern}, this minimal scenario is not realized in the large-$N$ limit of QCD$_3$. This will be remedied below by considering a more general scalar potential.

\subsubsection{$k  <{N_f \over 2}$}\label{sssec:smallk}

When~$k < {N_f \over 2}$, there are two proposed bosonic duals which together have been argued to cover the phase diagram of QCD$_3$. The analysis of each dual closely parallels the discussion in section~\ref{sssec:largek} above. The only essential difference is that the ranks of the dual gauge groups are now less than~$N_f$, so that we are in the regime discussed around~\eqref{eq:phismalln}:
\begin{itemize}
\item[1.)] As above, the first dual can be schematically described as follows,
\begin{equation}\label{eq:dualone}
U\left(\t N = k + {N_f \over 2} \leq N_f\right)_{\t k = - N} + \phi^{ai} + V_4\left(\phi\right) \qquad \left(a = 1, \ldots, \t N~;~i = 1, \ldots, N_f\right)~.
\end{equation}
Here~$V_4(\phi)$ is the quartic scalar potential in~\eqref{eq:quartpot}, and we again assume that~$\lambda_{1,2}$ are generic (subject to the various positivity bounds discussed above). This leads to the following phases as a function of~$\mu^2$:
\begin{itemize}
\item[1a.)] When~$\mu^2 > 0$ we are in a phase with~$\phi^{ai} = 0$, where all gauge and flavor symmetries are unbroken, and where all scalars are massive and can be integrated out. As in~\eqref{eq:irtqftone} this leaves the following TQFT in the deep IR,
\begin{equation}\label{eq:irtqftonebis}
U\left(k + {N_f \over 2 } \right)_{-N} \qquad \longleftrightarrow \qquad SU(N)_{k + {N_f \over 2 }}~.
\end{equation}
\item[1b.)] When~$\mu^2 < 0$ we find a phase where the vev of~$\phi^{ia}$ is described by~$\t N$ identical, non-vanishing singular values~$y_1 = \cdots = y_{\t N} \sim 1$ (see~\eqref{eq:phismalln} and~\eqref{eq:messmalln}), so that
\begin{equation}
\phi^{ai} \sim \left( \1_{\t N \times \t N}~,~\mathds{O}_{\t N \times \left(N_f - \t N\right)}\right)~, \quad {\CM_i}^j \sim \text{diag}\Big(\overbrace{\, 1 \, , \, \ldots \, , \, 1\,}^{\t N} \, , \,  \overbrace{\, 0 \, , \, \ldots \, , \, 0\, }^{N_f - \t N}\Big)~. 
\end{equation}
Therefore the~$U(\t N)$ gauge symmetry is completely higgsed, while the~$SU(N_f)$ flavor symmetry is broken as follows,
\begin{equation}\label{eq:symbrold}
SU(N_f) \; \rightarrow \; S\Big[U(\t N) \times U(N_f - \t N)\Big] = S\bigg[U\left({N_f \over 2} + k\right) \times U\left({N_f \over 2} - k\right)\bigg]~.
\end{equation}
The only massless fields in this phase are the NG bosons associated with this symmetry breaking, which parametrize the following Grassmannian target space, 
\begin{equation}\label{eq:grassrold}
\text{Gr}\left({N_f \over 2} + k, N_f\right) = {SU(N_f) \over S\bigg[U\left({N_f \over 2} + k\right) \times U\left({N_f \over 2} - k\right)\bigg]}~.
\end{equation}
\end{itemize}
\item[2.)] The second dual takes the following schematic form,
\begin{equation}\label{eq:dualtwo}
U\left(\hat N = {N_f \over 2} - k \leq N_f \right)_{\hat k = N} + \; {\hat \phi}^{a i} \, + \; \hat V_4(\hat \phi) \quad \left(a = 1, \ldots, \hat N~;~i = 1, \ldots, N_f\right)~.
\end{equation}
The quartic scalar potential takes the same form as in~\eqref{eq:quartpot}, with mass parameter~$\hat \mu$. This leads to the following phases as a function of~$\hat \mu^2$:
\begin{itemize}
\item[2a.)] When~$\hat \mu^2 > 0$ then~$\hat \phi^{ai} = 0$. All gauge and flavor symmetries are unbroken and, using level-rank duality the deep infrared is described by the following TQFT,
\begin{equation}
U\left({N_f \over 2} - k\right)_{N} \qquad \longleftrightarrow \qquad SU(N)_{k - {N_f \over 2}}~.
\end{equation}
\item[2b.)] When~$\hat \mu^2 < 0$ we find a phase with
\begin{equation}
\hat \phi^{ai} \sim \left(\1_{\hat N \times \hat N}~,~\mathds{O}_{\hat N \times \left(N_f-\hat N\right)}\right)~, \quad  {{\hat \CM}_i}^{\phantom{i} j} \sim \text{diag}\Big(\overbrace{\, 1 \, , \, \ldots \, , \, 1\,}^{\hat N} \, , \,  \overbrace{\, 0 \, , \, \ldots \, , \, 0\, }^{N_f - \hat N}\Big)~. 
\end{equation}
As in the discussion around~\eqref{eq:symbrold}, the~$U(\hat N)$ gauge symmetry is completely higgsed, while the flavor symmetry breaks according to the following pattern,
\begin{equation}
SU(N_f) \; \rightarrow \; S\Big[U(\hat N) \times U(N_f - \hat N)\Big] = S\bigg[U\left({N_f \over 2} - k\right) \times U\left({N_f \over 2} + k\right)\bigg]~.
\end{equation}
The target space for the massless NG bosons is therefore the same as in~\eqref{eq:grassrold}, since $\text{Gr}\left({N_f \over 2} - k, N_f\right) = \text{Gr}\left({N_f \over 2} + k, N_f\right)$. All other particles are massive.

\end{itemize}

\end{itemize}

The phases in~1a.)~and~2a.)~above coincide with the asymptotic phases of QCD$_3$ for large (positive or negative) flavor-singlet mass~$m$. In these regimes we can therefore identify~$\mu^2 \sim m \Lambda~(m \gg \Lambda)$ and~$\hat \mu^2 \sim - m \Lambda~(m \ll - \Lambda)$. The phases in~1b.)~and~2b.)~are identical at low energies. It was proposed in~\cite{Komargodski:2017keh} that these three distinct phases coincide with the phases of QCD$_3$ as a function of its flavor-singlet mass parameter~$m$. 
As before, we see that the bosonic dual theories, furnished with only a quartic scalar potential, are not able to capture the large-$N$ dynamics that we have uncovered in QCD$_3$. 

\subsection{A Dual Scalar Potential for Large-$N$ QCD$_3$}\label{ssec:dualsp}

The aim of this subsection is to propose a scalar potential such that the dynamics of the bosonic dual theories agrees with that of large-$N$ QCD$_3$. As we have seen above, it is necessary to include terms beyond the quartic order. 

\subsubsection{ $k\geq N_f/2$}\label{sssec:largeknew}

As we saw in section~\ref{sssec:largek}, a generic quartic potential for the bosonic dual theory suggests a single transition between the two asymptotic phases with~$SU(N)_{k \pm {N_f \over 2}}$ Chern-Simons theories. Essentially, this is because the eigenvalues of the meson matrix~${\CM_i}^j = \text{diag}\left(y_1^2, \ldots, y_{N_f}^2\right)$ make a single transition from a vacuum where all~$y_i = 0$ to a vacuum where all~$y_i$ are equal but non-zero. In order to reproduce the~$N_f+1$ different large-$N$ vacua of~QCD$_3$ uncovered in section~\ref{ssec:disksb}, we must allow the individual eigenvalues~$y_i$ to make the transition one eigenvalue at a time. The structure of the required potential for the bosonic operator~${\CM_i}^j  = \phi^\dagger_{i} \phi^{j}$ can be inferred by comparing to the effective potential~\eqref{eq:veffviafx} for its fermionic dual~${M_i}^j = {1 \over N} \b \psi_i \psi^j$. The salient features of this effective potential are its single-trace structure, with two exactly degenerate minima for each eigenvalue~$x_i$ (this allows each~$x_i$ to transition independently), and the fact that the potential is~$\CO(N)$ in the large-$N$ limit. 

\begin{figure}[!th]
\centerline{\includegraphics[scale=1]{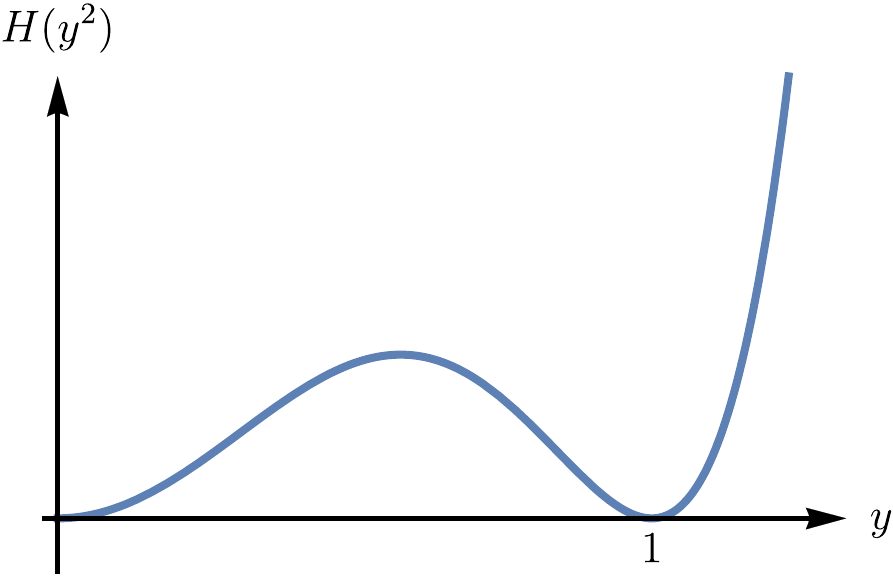}}
\caption{The function $H(y^2)$ entering in \eqref{eq:bosmesstpot}, with two degenerate minima at $y=0,1$.}
\label{PlotH}
\end{figure}

Motivated by these properties, we propose that the effective potential for the dual bosonic meson~${\CM_i}^j$ takes the following form,
\begin{equation}\label{eq:bosmesstpot}
V\left({ \CM}\right) = N \Lambda^3 \sum_{i = 1}^{N_f} H(y_i^2)~.
\end{equation}
Here~$H(y^2)$ is a function that does not depend on~$N$ and has two exactly degenerate minima. After a suitable rescaling, we can place them at~$y  = 0$ and~$y = 1$. See figure \ref{PlotH} for an example of such a function. 

Before spelling out why this form of~$H(y^2)$ leads to the desired properties, several comments are in order:
\begin{itemize}
\item The properties required of~$H(y^2)$ are impossible to achieve using a quartic polynomial. This means that the quartic potentials reviewed in section~\ref{ssec:oldpot} (and assumed in most of the recent literature) cannot provide a correct dual description for the vacuum structure of large-$N$ QCD$_3$. 
\item While any function~$H(y^2)$ satisfying the properties listed above will lead to the correct vacuum structure, it is possible to satisfy all requirements by choosing~$H(y^2)$ to be a sextic polynomial,
\begin{equation}\label{eq:sextich}
H(y^2) \sim y^2 (y^2 -1)^2~.
\end{equation}
Choosing such an~$H(y^2)$ has the interesting feature of making the bosonic dual theory renormalizable. Boson-fermion dualities with sextic scalar potentials were also recently considered in~\cite{Aharony:2018pjn}, albeit for different reasons and in a different parameter regime (large~$N, k$  and fixed~$N \over k$). In order to simplify the presentation below, we will assume that~$H(y^2)$ has been chosen as in~\eqref{eq:sextich}. 
\item With the scalar potential~\eqref{eq:bosmesstpot}, the entire bosonic dual Lagrangian~\eqref{eq:duallag} scales like~$N$,
\begin{equation}\label{eq:duallagbis}
\SL = N \left((D^\mu \phi)^\dagger_i D_\mu \phi^i \pm {1 \over 4 \pi} \Tr \left(C \wedge dC + {2 \over 3} C \wedge C \wedge C\right) + \Lambda^3 \tr\left(H\left({\CM_i}^j = \phi_i^\dagger \phi^j\right)\right)\right)~.
\end{equation}
This makes it clear that the large-$N$ limit is nothing but a conventional weak-coupling limit in the bosonic dual theory. 

\item The two degenerate minima of the function~$H(y^2)$ in~\eqref{eq:sextich} are not accounted for by a symmetry, so that this degeneracy can only be achieved by fine-tuning. Consequently, we do not expect it to be respected by quantum corrections. Fortunately, we will only need to insist on this fine-tuning at leading order in the large-$N$ (or semiclassical) limit, while quantum corrections are suppressed by~$1 \over N$. 

\end{itemize}

\begin{figure}[!th]
\centerline{\includegraphics[scale=0.48]{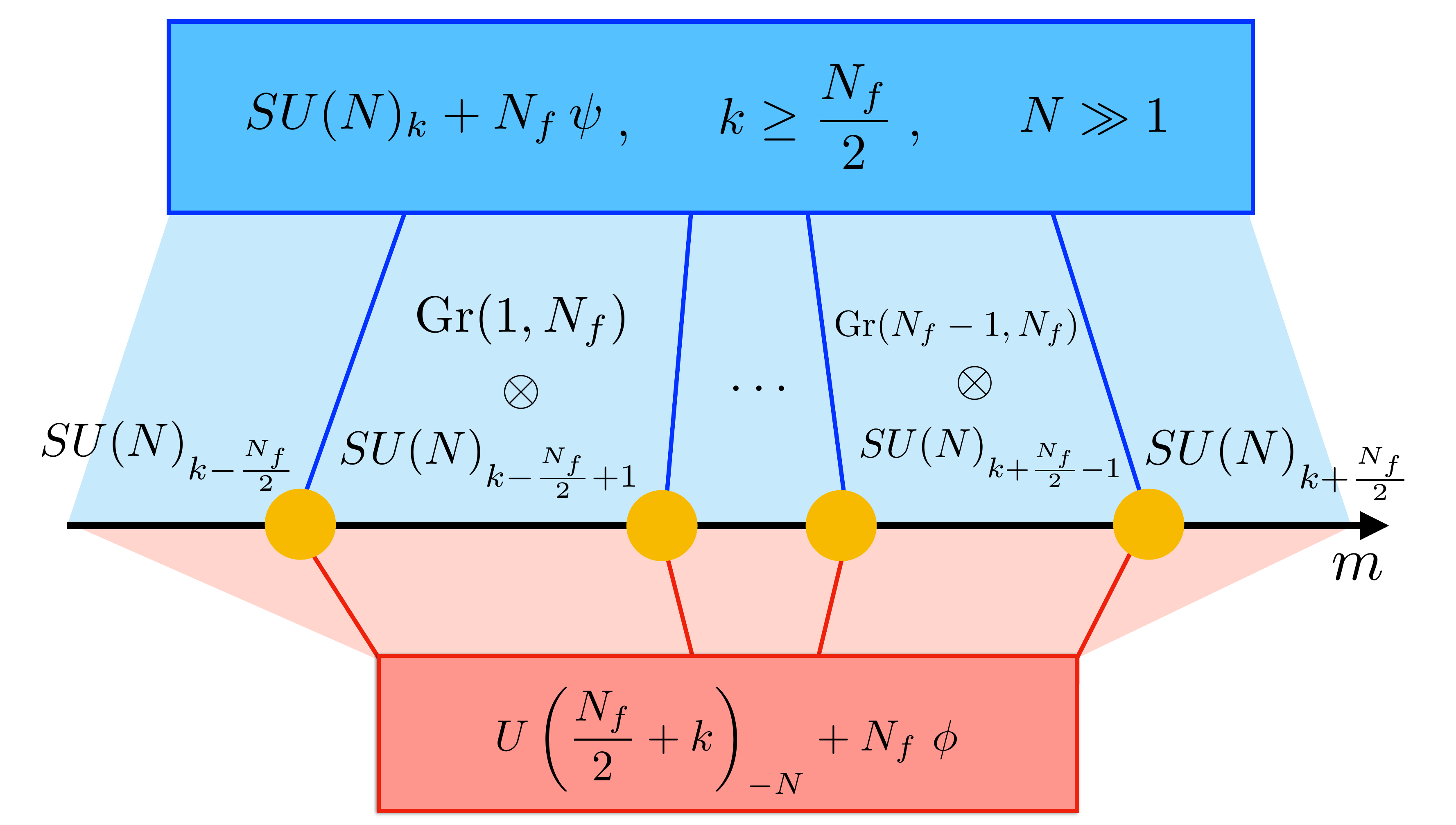}}
\caption{Phases of large-$N$ QCD$_3$ (shown in blue) with~$k \geq {N_f\over 2}$ as a function of the flavor-singlet mass~$m$. There are~$N_f+1$ phases, which generically contain both NG bosons (with Grassmannian target spaces) associated with different patterns of flavor-symmetry breaking, as well as a Chern-Simons TQFT in the deep IR. There are~$N_f$ first-order phase transitions indicated by yellow dots. The same phases and transitions are captured by the bosonic dual (shown in red), if the scalar potential for~$\phi$ takes a suitable form.}
\label{phasesLarge}
\end{figure}

In order to analyze the consequences of the scalar potential~\eqref{eq:bosmesstpot}, we use an~$SU(N_f)$ Weyl transformation to order the~$y_i$ in descending order as in~\eqref{eq:phibign}, i.e.~$y_1 \geq y_2 \geq \cdots \geq y_{N_f} \geq 0$. Together with the fact that~$H(y^2)$ in~\eqref{eq:sextich} has two exactly degenerate minima at~$y = 0$ and~$y = 1$, it follows that there are~$N_f + 1$ independent vacuum sectors that are not related by~$SU(N_f)$ flavor transformations,
\begin{equation}\label{eq:ypvac}
y_1 = \cdots = y_{N_f- p} = 1~, \qquad y_{N_f - p+1} = \cdots = y_{N_f} = 0~, \qquad p \in \{0, 1, \ldots, N_f\}~.
\end{equation}
Here the integer~$p$ counts the number of eigenvalues in the~$y = 0$ vacuum. Comparing with~\eqref{eq:phibign}, we see that the~$U\left(k + {N_f \over 2}\right)$ gauge symmetry is higgsed to~$U\left(k - {N_f \over 2} + p\right)$. Similarly, the~$SU(N_f)$ flavor symmetry is broken to~$S \left(U(N_f - p) \times U(p)\right)$. The only gapless degrees of freedom that remain in each vacuum are the NG bosons resulting from this symmetry breaking. They parametrize a Grassmannian target space,
\begin{equation}\label{eq:grasmanbis}
\text{Gr}(p, N_f) = {SU(N_f) \over S\left(U(N_f - p) \times U(p)\right)} = {U(N_f) \over U(p) \times U(N_f -p)}~,
\end{equation}
as in~\eqref{eq:grassdef}. Similarly, the unbroken gauge symmetry leads to a~$U\left(k - {N_f \over 2} + p\right)_{-N}$ Chern-Simons theory, which is level-rank dual to~$SU(N)_{k - {N_f \over 2} + p}$. Together with the NG bosons in~\eqref{eq:grasmanbis}, this exactly matches the low-energy degrees of freedom in~\eqref{eq:largenefts} (see also figure~\ref{phasesLarge}). In each vacuum, the mapping between the bosonic meson eigenvalues~$y^2_i$ and the fermionic meson eigenvalues~$x_i$ is as follows,
\begin{equation}\label{eq:xymapping}
y^2_{1, \ldots, N_f - p} = \half\left(1 + x_{p+1, \ldots, N_f}\right)~, \qquad y^2_{N_f - p +1, \ldots, N_f} = \half \left(1 + x_{1, \ldots, p}\right)~.
\end{equation}

At low energies, the diagonal abelian gauge field~$\Tr C$ constructed using the microscopic~$U\left(k - {N_f \over 2}\right)$ gauge field~$C$ flows to a linear combination of the diagonal abelian gauge field (still denoted~$\Tr C$) constructed using the unbroken~$U\left(k - {N_f \over 2} + p\right)$ gauge field in the deep IR, and the K\"ahler connection of the Grassmannian NG target space. Thus, the microscopic baryon number~\eqref{eq:dualmicroB} flows to the following expression in the deep IR,
\begin{equation}\label{BaryonS}
B = {1 \over 2 \pi}\int_{\CM_2} \Tr dC \quad \longrightarrow \quad B_\text{IR} = {1 \over 2 \pi} \int_{\CM_2} \omega + {1 \over 2 \pi}\int_{\CM_2} \Tr dC~.
\end{equation}
Here~$\omega$ is the K\"ahler form of the Grassmannian manifold, normalized so that its periods are properly quantized and~$B \in \Z$. Its contribution to~$B$ in~\eqref{BaryonS} is the standard skyrmion contribution to baryon number. By contrast, the second term in~\eqref{BaryonS} is the monopole charge of the~$U\left(k - {N_f \over 2} + p\right)_{-N}$ Chern-Simons theory in the deep IR. Even though it vanishes as an operator, it can still couple to the~$U(1)_B$ background gauge field (see for instance~\cite{Closset:2012vg,Closset:2012vp} for a detailed explanation with examples). The fact that baryon number receives contributions  from both skyrmions and Chern-Simons anyons reflects the fact (mentioned at the end of section~\ref{ssec:vacefts}) that gauge-neutral baryons, described by conventional skyrmions, can in principle decay into deconfined anyonic quarks, whose worldlines are captured by the Wilson lines of the low-energy Chern-Simons TQFT. Note however that the decay process itself involves the microscopic degrees of freedom of the theory and is not captured by the universal low-energy physics.

It is straightforward to check that small mass deformations also work exactly as in the fermionic case. Such a deformation to the scalar potential takes the following form,
\begin{equation}\label{eq:bosmassdef}
\Delta V(\CM) = N \tr\left(\mu^2 \, \CM \right)~.
\end{equation}
On the minima of the scalar potential of the massless theory, $\CM$ has eigenvalues~$0$ or~$1$, and hence it satisfies the constraint~$\CM(\CM-\1) = 0$. Implementing this with a Lagrange multiplier, as in~\eqref{eq:lagmmulmin}, leads to the vacuum alignement condition~$[{ \mu^2}, \CM] = 0$, so that we can simultaneously diagonalize~${ \mu^2} = \text{diag}\left(\mu^2_1, \ldots, \mu^2_{N_f}\right)$ and~$\CM = \Lambda \text{diag}\left( y_1^2, \ldots, y_{N_f}^2\right)$. Note that each mass eigenvalues~$\mu_i^2$ can take arbitrary real values. The mass deformation~\eqref{eq:bosmassdef} then reduces to
\begin{equation}
\Delta V(y_i) = N \Lambda \sum_{i = 1}^{N_f} \mu^2_i y_i^2~.
\end{equation}
Ordering the mass eigenvalues in ascending order,
\begin{equation}
\mu_1^2 \leq \cdots \leq \mu_{N_f-p}^2 \leq 0 \leq \mu_{N_f- p+1}^2 \leq \cdots \mu_{N_f}^2~,
\end{equation}
we find that the first~$N_f-p$ eigenvalues are at~$y_i = 1$, while the remaining~$p$ eigenvalues are at~$y_i = 0$, exactly as in~\eqref{eq:ypvac}. However, the Grassmannian NG manifolds in each superselection sectors are lifted due to the vacuum alignment condition. Comparing with~\eqref{eq:pposms}, we can establish the following map between the mass eigenvalues,
\begin{equation}
\mu_{1, \ldots, N_f - p}^2 \sim \Lambda m_{p+1, \ldots, N_f}~, \qquad \mu^2_{N_f - p + 1, \ldots, N_f} \sim \Lambda m_{1, \ldots, p}~,
\end{equation}
where the proportionality constant is the same in both relations to ensure~$SU(N_f)$ covariance. Recall that the zero-mass point of QCD$_3$ is meaningful at leading order in the large-$N$ limit. In the bosonic dual this point corresponds to fine-tuning the scalar potential to achieve precisely degenerate vacua at~$y  = 0,1$, which is also meaningful because the large-$N$ limit of the bosonic theory is semiclassical. 

\subsubsection{$k<N_f/2$}\label{sssec:smallknew}

It is straightforward to repeat the discussion of the preceding section for the two bosonic duals in section~\ref{sssec:smallk}. If we work in the bosonic dual~\eqref{eq:dualone} with gauge group $U\left(\t N = k + {N_f \over 2}\right)$, the bosonic meson operator~$\CM = \Lambda \, \text{diag}\left(y_1^2, \ldots, y^2_{k + {N_f \over 2}}, 0, \ldots, 0\right)$ has rank at most~$k + {N_f \over 2}$. If we add the scalar potential~\eqref{eq:bosmesstpot}, only the first~$k + {N_f \over 2}$ terms in the sum will be nonvanishing. This leads to~$k + {N_f \over 2} + 1$ vacua, labeled by an integer~$q$,
\begin{equation}
y_1 = \cdots = y_{k + {N_f \over 2} - q} = 1~, \qquad y_{k + {N_f \over 2} - q + 1} = \cdots = y_{k + {N_f \over 2}} = 0~, \qquad q \in \left\{0, 1, \ldots, k + {N_f \over 2}\right\}~.
\end{equation}
In other words, $q$ counts the number of vanishing~$y_i$.  Comparing with~\eqref{eq:phismalln} (with $\t N = k + {N_f \over 2}$) we find that the~$U\left(k + {N_f \over 2}\right)$ gauge symmetry is higgsed to~$U(q)$, while the global~$SU(N_f)$ flavor symmetry is broken to~$S \left(U\left(k + {N_f \over 2} - q\right) \times U\left({N_f \over 2} - k + q\right)\right)$. Using level-rank duality, we can summarize the low-energy degrees of freedom coming from the NG bosons and the Chern-Simons theory as follows,
\begin{equation}
\text{Gr}\left({N_f \over 2} - k + q \, , N_f\right) \; \otimes \; SU(N)_q~, \qquad q \in \left\{0, 1, \ldots, k + {N_f \over 2}\right\}~.
\end{equation}
Comparing with~\eqref{eq:largenefts}, we see that the~$k + {N_f \over 2} + 1$ phases covered by~$q$ correspond to choosing~$p = {N_f \over 2} - k + q$ in that formula, i.e.~we cover only the phases labeled by
\begin{equation}\label{eq:pcovone}
p \in \left\{ {N_f \over 2} - k, \ldots, N_f\right\}~.
\end{equation}
Note that~$q = 0$ labels the phase with only NG bosons and no TQFT. 

If we repeat this analysis for the second bosonic dual~\eqref{eq:dualtwo}, with gauge group $U\left(\hat N = {N_f\over 2} - k\right)$, we similarly obtain the following superselection sectors, 
\begin{equation}
\text{Gr}\left({N_f \over 2} - k - \hat q \, , N_f\right) \; \otimes \; SU(N)_{-\hat q}~, \qquad \hat q \in \left\{0, 1, \ldots, {N_f \over 2} - k\right\}~.
\end{equation}
Comparing with~\eqref{eq:largenefts}, we see that the~${N_f \over 2} - k + 1$ phases covered by~$\hat q$ correspond to choosing~$p = {N_f \over 2} - k- \hat q$ in that formula, so that we cover the phases labeled by
\begin{equation}\label{eq:pcovtwo}
p \in \left\{0, 1, \ldots, {N_f \over 2} - k\right\}~.
\end{equation}

\begin{figure}[!th]
\centerline{\includegraphics[scale=0.48]{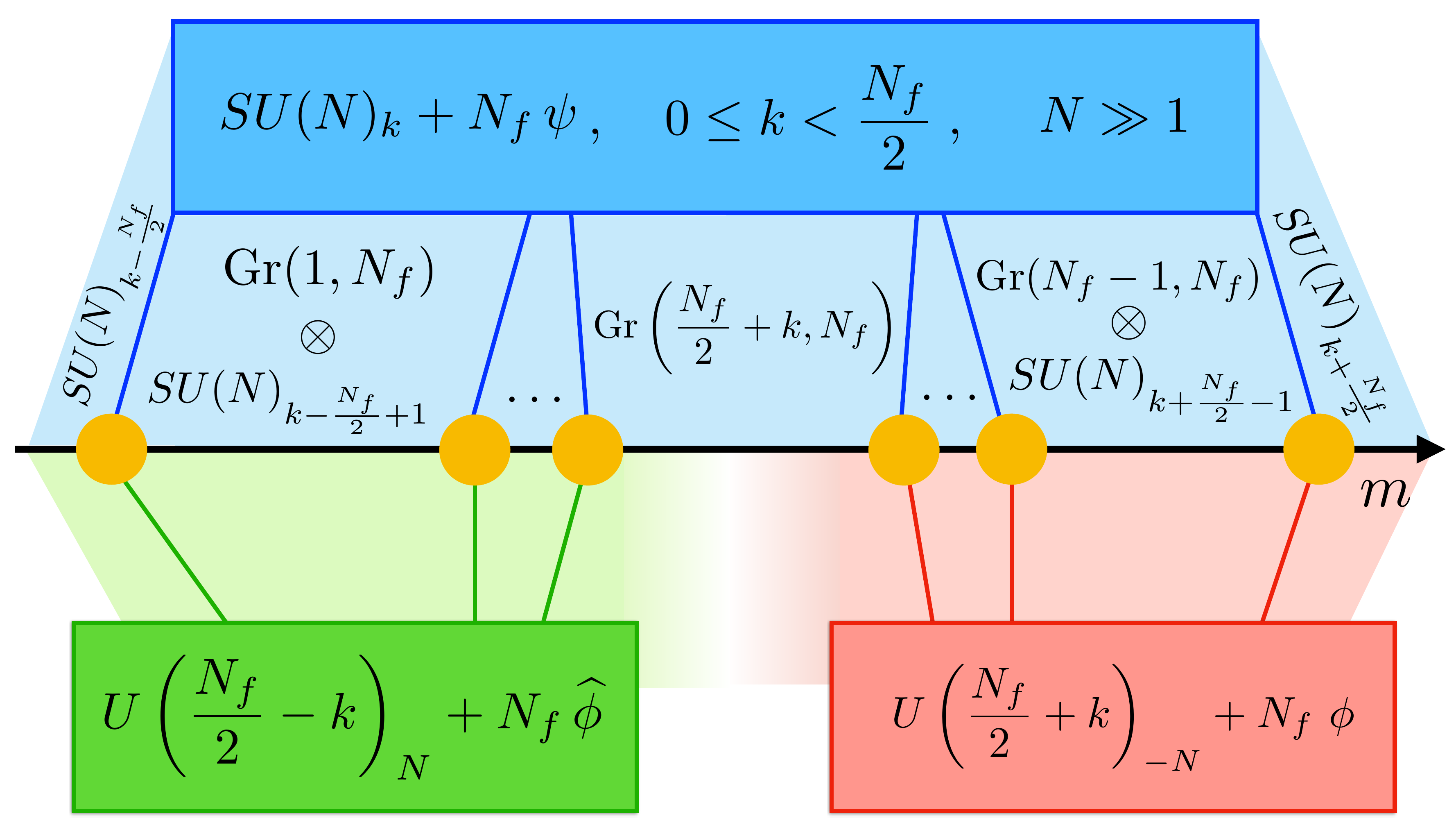}}
\caption{Phases of large-$N$ QCD$_3$ (shown in blue) with~$0 \leq k < {N_f\over 2}$ as a function of the flavor-singlet mass~$m$. There are~$N_f+1$ phases, which generically contain both NG bosons (with Grassmannian target spaces) associated with different patterns of flavor-symmetry breaking, as well as a Chern-Simons TQFT in the deep IR. There are~$N_f$ first-order phase transitions indicated by yellow dots. The same phases and transitions are captured by the two mutually non-local bosonic dual theories (shown in green and red), if their respective scalar potentials are chosen appropriately.}
\label{phases}
\end{figure}

If we now compare the values of~$p$ in~\eqref{eq:pcovone} and~\eqref{eq:pcovtwo} that are covered by the two bosonic duals, we see that every phase is covered by at least one of the duals. Moreover, the phase with~$p = {N_f \over 2}- k$, which only has NG bosons and no TQFT, lies in the common regime of validity of both duals.  This is the only vacuum sector that can be simultaneously described in both duals. The phase diagram and the regime of validity of the two bosonic duals are summarized in figure~\ref{phases}.

Note that as in the discussion around~\eqref{BaryonS}, the~$U(1)_B$ baryon charge~$B$ in the various sectors generically receives contributions from both skyrmions associated with the NG boson sigma model, as well as from the Chern-Simons TQFT. 

\subsubsection{Beyond Leading Order in the Large-$N$ Expansion}\label{sssec:subleaddual}

Having understood how to describe the exactly degenerate vacua of the large-$N$ theory by choosing suitable scalar potentials for the bosonic duals, we now discuss how to modify the duals to describe the phenomena uncovered in section~\ref{sec:oneovern}, which are subleading in the large-$N$ expansion. For simplicity we focus on the case~$k \geq {N_f \over2}$, for which there is a single bosonic dual. The case with two mutually non-local bosonic duals can be analyzed along similar lines.

In QCD$_3$ at large~$N$, the~$\CO(1)$ correction to the effective potential is given by~\eqref{eq:antiferro}, which we repeat here,
\begin{equation}\label{eq:fermantiferr}
V(x_i)\big|_{\CO(1)} = \Lambda^3 \Delta \sum_{i, j = 1}^{N_f} x_i x_j~, \qquad \Delta > 0~.
\end{equation}
Recall that the sign of~$\Delta$ was fixed by appealing to the Vafa-Witten theorem~\cite{Vafa:1983tf}. It is straightforward to replicate this effective potential in the bosonic dual theory by using the mapping~\eqref{eq:xymapping}, which is valid in the vacua of the large-$N$ theory. Substituting into~\eqref{eq:fermantiferr}, we find 
\begin{equation}\label{eq:xyferromag}
 \Lambda^3 \Delta \sum_{i, j = 1}^{N_f} x_i x_j = \Lambda^3 \Delta \sum_{i, j =1}^{N_f} \left(2 y_i^2 - 1\right)\left(2 y_j^2 - 1\right)~.
\end{equation}
Multiplying this out, we find an immaterial constant shift of the energy, as well as an expected additive shift~$\Delta \mu^2 \sim \Lambda^2 \Delta$ in the flavor-singlet mass~$\mu^2$. The essential new term that first appears at~$\CO(1)$ in the large-$N$ expansion is the following double-trace coupling,
\begin{equation}\label{eq:bosdoubtr}
V(\CM)\big|_{\CO(1)} = 4 \Lambda^3 \Delta \sum_{i, j = 1}^{N_f} y_i^2 y_j^2 = 4 \Lambda^3 \Delta  \tr\left(\CM\right)^2~, \qquad \Delta > 0~.
\end{equation}
Given that the leading large-$N$ vacua are correctly captured in the fermionic~$x$-description and the bosonic~$y$-description, the equality~\eqref{eq:xyferromag} guarantees that the effects appearing at~$\CO(1)$ in the large-$N$ expansion as described in section~\ref{sec:oneovern} are correctly captured by adding the~$\CO(1)$ double trace potential~\eqref{eq:bosdoubtr} to the bosonic dual theory~\eqref{eq:duallagbis}. 

\begin{figure}[!th]
\centerline{\includegraphics[scale=0.55]{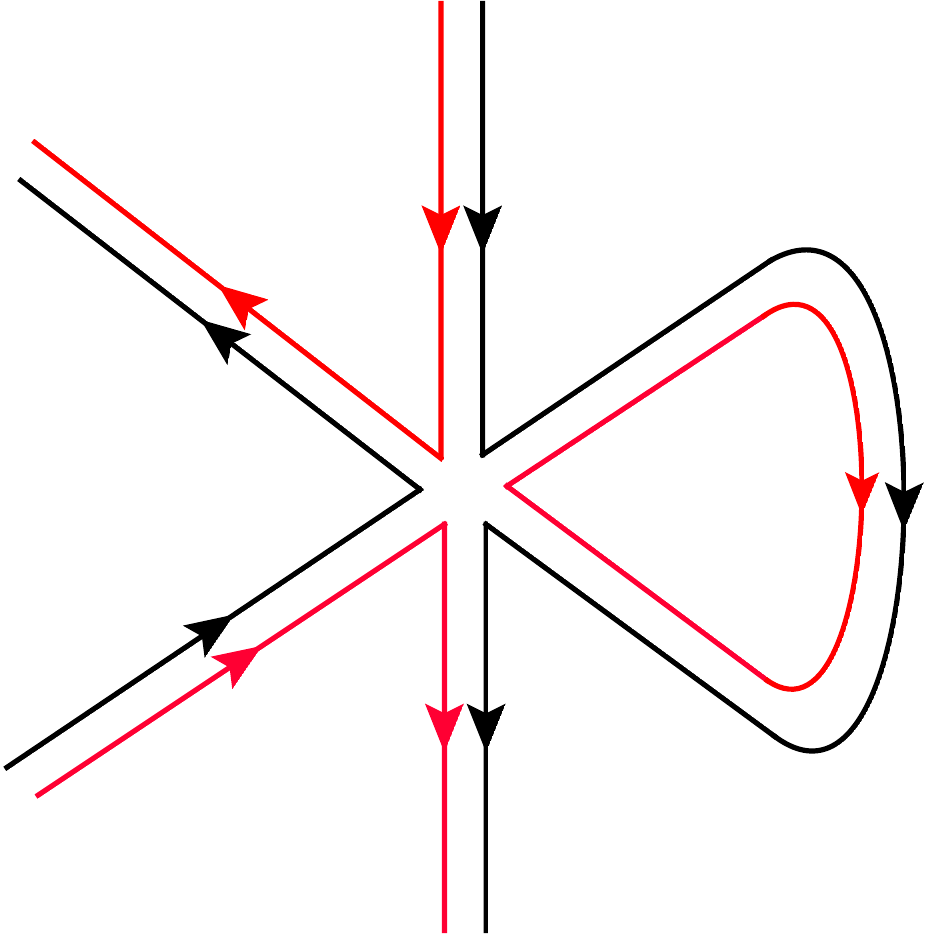} \hspace{.5cm}\includegraphics[scale=0.55]{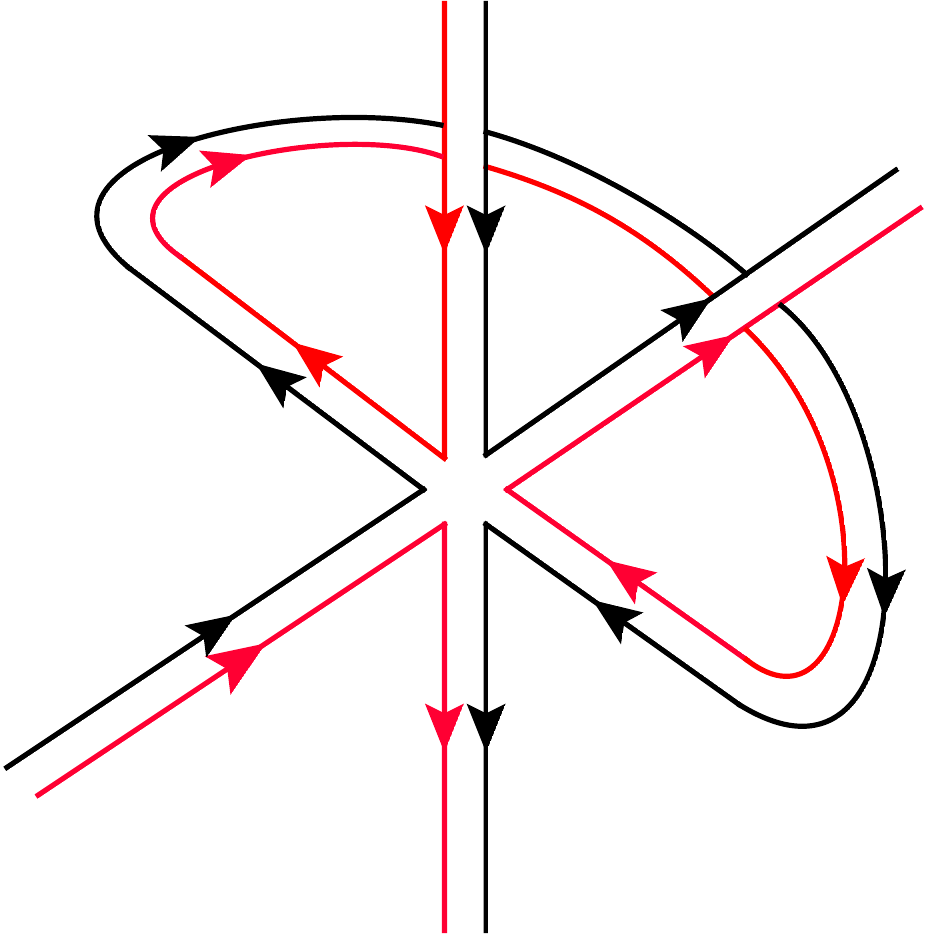}}
\caption{Different contractions of $\phi \bar \phi$ pairs can lead to single- or multi-trace vertices. Here red lines represent flavor indices and black lines indicate gauge indices.}
\label{doubleline}
\end{figure}

So far we have treated the bosonic dual theories classically and engineered their scalar potentials to correctly reproduce the behavior of large-$N$ QCD$_3$. We must now discuss to what extent these choices are compatible with quantum effects, e.g.~to what extent the scalar potentials are renormalized in the quantum theory. For the purposes of this discussion it is again convenient to choose the single-trace potential as in~\eqref{eq:bosmesstpot} and~\eqref{eq:sextich}, with a leading sextic term~$\sim N \tr\left(\CM^3\right) = N \tr \left( \left(\b \phi \phi\right)^3\right)$. This interaction leads to a vertex with three~$\phi$ and three~$\b \phi$ fields. Contracting a~$\phi \b \phi$ pair can either lead to a single-trace quartic coupling~$\sim \tr \left(\CM^2\right)$ or to a double-trace quartic~$\sim \tr\left(\CM\right)^2$ as in~\eqref{eq:bosdoubtr} (see figure~\ref{doubleline}). In three dimensions, such a contraction is linearly divergent,~since $\int {d^3 p \over p^2} \sim \Lambda_\text{UV}$, where~$\Lambda_\text{UV}$ is a UV cutoff.  

\noindent Similarly, contracting two~$\phi \b \phi$ pairs leads to a quadratically divergent mass renormalization~$\sim \Lambda_\text{UV}^2$. These power-law divergences require appropriate counterterms that must be fine-tuned to achieve a scalar potential whose structure is as described above. In particular, this means that the coefficients of these terms in the potential are not calculable, even though the bosonic dual theory is weakly coupled and can even be renormalizable. 

For instance, the~$\CO(N)$ scalar potential must be fine-tuned to the single-trace form in~\eqref{eq:bosmesstpot} (with~$H(y^2) \sim y^2 (y^2-1)^2$ as in~\eqref{eq:sextich}). However, this fine-tuning is not sufficient to guarantee that the coefficient~$\Delta$ of the double-trace potential~\eqref{eq:bosdoubtr} is positive. Ensuring this can be viewed as a further tuning, which is required to achieve the desired phase structure at~$\CO(1)$ in the large-$N$ expansion. It is however suggestive that a one-loop contraction of the single-trace sextic~$\sim N \tr \left( \left(\b \phi \phi\right)^3\right)$ with propagator~$\langle \phi \b \phi\rangle \sim{1 \over N}$ precisely induces an~$\CO(1)$ double-trace term of the form~\eqref{eq:bosdoubtr}, with a positive but formally UV-divergent value of~$\Delta$. If we remove the divergence using a double-trace counterterm, the positivity of~$\Delta$ is no longer automatic and must be ensured by tuning.

\bigskip
\bigskip\bigskip

\begin{center}
\large\textbf{Acknowledgments}
\end{center}
\vspace{-3pt}

\noindent We would like to thank O.\,Aharony, S.\,Caron-Huot,  N.\,Seiberg, S.\,Sugimoto, and C.\,Vafa for discussions. We also thank the authors of~\cite{Argurio:2019tvw} for sharing their results prior to publication, and for comments on a draft of our paper. A.\,A, T.\,D., and G.\,F. would like to thank the Simons Center for Geometry and Physics for hospitality during the initial stages of this project. The work of A.A. has been supported by STFC grant ST/P00055X/1. T.D. is supported by the Mani L. Bhaumik Presidential Chair in Theoretical Physics at UCLA. G.F. is supported by ERC STG grant 639220 and by the Swedish Research Council under grant 2018-05572.  Z.K. is supported in part by Simons Foundation grant 488657 (Simons Collaboration on the Non-Perturbative Bootstrap).  Any opinions, findings, and conclusions or recommendations expressed in this material are those of the authors and do not necessarily reflect the views of the funding agencies.

\newpage


\bigskip

\renewcommand\refname{\bfseries\large\centering References\\ \vspace{-0.4cm}
\addcontentsline{toc}{section}{References}}

\bibliographystyle{utphys.bst}
\bibliography{largenqcd3.bib}
	
\end{document}